\documentclass[aps, prd, floats, floatfix, onecolumn, superscriptaddress, nofootinbib, showpacs, letterpaper, 10pt, preprintnumbers, notitlepage]{revtex4-1}

\usepackage{graphicx}
\usepackage{amsmath}
\usepackage{amsfonts}
\usepackage{amssymb}
\usepackage{amstext}
\usepackage{tensor}
\usepackage[english]{babel}
\usepackage{blindtext}
\usepackage{helvet}
\usepackage{microtype}
\usepackage{bbm}
\usepackage{fullpage}
\usepackage{dsfont}
\usepackage[pdftex, pdftitle={Number Density of Peaks in a Chi-Squared Field},
pdfauthor={Jolyon K. Bloomfield, Stephen H. P. Face, Alan H. Guth, Saarik Kalia, Casey Lam, Zander Moss},
bookmarks,bookmarksopen=false,
pdfstartview={FitH},
linktocpage=true
]{hyperref}

\newcommand{\prob}{\mathrm{P}}
\newcommand{\N}{{\mathcal N}}
\newcommand{\mat}[1]{\begin{bmatrix}#1\end{bmatrix}}
\newcommand{\mydet}[1]{\left|\det \left(#1\right)\right|}
\newcommand{\tr}{\operatorname{Tr}}

\begin{document}

\title{Number Density of Peaks in a Chi-Squared Field}

\author{Jolyon K. Bloomfield}
\email{jolyon@mit.edu}
\author{Stephen H. P. Face}
\email{face@mit.edu}
\author{Alan H. Guth}
\email{guth@ctp.mit.edu}
\author{Saarik Kalia}
\email{skalia@mit.edu}
\author{Casey Lam}
\email{caseylam@mit.edu}
\author{Zander Moss}
\email{zander@mit.edu}
\affiliation{Center for Theoretical Physics, Laboratory for Nuclear Science, and Department of Physics, Massachusetts Institute of Technology, Cambridge, MA 02139, USA}

\date{\today}

\preprint{MIT-CTP/4867}

\begin{abstract}
We investigate the statistics of stationary points in the sum of squares of $N$ Gaussian random fields, which we call a ``chi-squared'' field. The behavior of such a field at a point is investigated, with particular attention paid to the formation of topological defects. An integral to compute the number density of stationary points at a given field amplitude is constructed. We compute exact expressions for the integral in various limits and provide code to evaluate it numerically in the general case. We investigate the dependence of the number density of stationary points on the field amplitude, number of fields, and power spectrum of the individual Gaussian random fields. This work parallels the work of Bardeen, Bond, Kaiser and Szalay, who investigated the statistics of peaks of Gaussian random fields. A number of results for integrating over matrices are presented in appendices.
\end{abstract}


\maketitle

\section{Introduction}

Thirty years ago, Bardeen, Bond, Kaiser and Szalay (BBKS) published ``The statistics of peaks of Gaussian random fields'' \cite{Bardeen1986}, wherein they undertook a tour-de-force computation of everything they could analyze regarding peaks in Gaussian random fields. In this paper, we undertake the equivalent research program for fields with a different statistical nature: fields that are the sum of squares of Gaussian random fields, which we dub chi-squared fields after the statistical distribution.

While not nearly so ubiquitous as Gaussian random fields, the sum of squares of Gaussian random fields naturally arises whenever a $SO(N)$ symmetry exists in field space. Our particular motivation for studying such fields arises from a model of hybrid inflation \cite{Linde1993,Lyth2012,Guth2012}, where $N$ ``waterfall fields'' $\phi_\alpha$ each develop Gaussian quantum fluctuations, resulting in density perturbations dependent upon the sum of their squares, $\sum_{\alpha} \phi_\alpha^2$. We are interested in understanding the number density of black holes that form in such theories, which requires understanding the number density of large peaks that could then undergo gravitational collapse.

The approach in this paper closely mirrors the approach taken by BBKS and involves constructing an expression to count the number of peaks in a field realization before taking the ensemble average to compute the number density of stationary points at a given field amplitude. Unlike BBKS, we draw heavily from multivariate statistical theory, which we use to analytically integrate over many more fields than is necessary for the Gaussian case. We eventually write the number density for stationary points in terms of a 9-dimensional integral that must be numerically integrated. Python code to compute these integrals has been made publicly available.

Our final result depends on the number of fields $N$, the dimensionless field amplitude, and three moments of the Gaussian power spectrum. Ratios of these moments form a dimensionless number that describe the spread of the power spectrum, ranging from a delta function to a slowly-decaying tail. Only $N$, the field amplitude and this ratio contribute to the numerical integral, with the moments of the power spectrum providing dimensional scaling for the number density. We investigate how the number density of stationary points depends on this dimensionless ratio, the field amplitude, the number of fields, and the type of stationary point in question. In some limits, we are able to compute analytic estimates for the number densities.

We begin with a discussion of the structure of chi-squared fields in Section \ref{sec:structure}, before computing an integral expression for the number density of stationary points in Section \ref{sec:nfields}. We prepare for performing numeric integration in Section \ref{sec:numintegrate}, and present numerical results in Section \ref{sec:numerics} before concluding in Section \ref{sec:conclusions}. A number of technical appendices present details on multivariate statistics and matrix integration.

\section{The Structure of Chi-Squared Fields} \label{sec:structure}

Consider a field
\begin{align}
\Phi = \sum_{\alpha = 1}^N \phi_\alpha^2
\end{align}
where $\phi_\alpha$ are $N$ independent centered Gaussian random fields with identical power spectra. The expectation value of $\Phi$ at a point is given by
\begin{align}
\langle \Phi \rangle = N \langle \phi^2 \rangle = N \sigma_0^2
\end{align}
where the variance $\langle \phi^2 \rangle = \sigma^2_0$ is defined in Appendix \ref{app:stats}, and is the same for each $\phi_\alpha$. Letting $\bar{\nu}^2 = \Phi / \sigma_0^2$ with $\bar{\nu} > 0$, we have that $\bar{\nu}^2$ is distributed according to a $\chi^2(N)$ distribution, $\bar{\nu}^2 \sim \chi^2(N)$. Hence, the probability density $dP/d(\bar{\nu}^2)$ is given by
\begin{align}
\frac{dP}{d(\bar{\nu}^2)} = \frac{\bar{\nu}^{N-2} e^{-\bar{\nu}^2/2}}{2^{N/2} \Gamma(N/2)}
\end{align}
which peaks at $\bar{\nu}^2 = N - 2$. We can also write this as
\begin{align}\label{eq:dpdnu}
\frac{dP}{d\bar{\nu}} = \begin{cases}
0 & \text{for} \ \bar{\nu} < 0 \\
\displaystyle\frac{\bar{\nu}^{N-1} e^{-\bar{\nu}^2/2}}{2^{N/2 - 1} \Gamma(N/2)} & \text{for} \ \bar{\nu} > 0.
\end{cases}
\end{align}
The peak of this distribution lies at $\bar{\nu} = \sqrt{N-1}$.

Let us now think about the characteristics of $\Phi$ beyond its distribution at a point, and look at the three-dimensional features of the field\footnote{Throughout this paper, we work in 3D space. Generalizations to other dimensions are for the most part straightforward.}. When $N = 1$, a single field $\phi$, centered around zero, will have regions of positive and negative $\phi$, surrounded by two-dimensional surfaces where $\phi = 0$. The field $\phi^2$ will then have regions of positive $\phi^2$, surrounded by the same two-dimensional surfaces. These are effectively domain walls in $\Phi$, and they are ubiquitous.

When $N = 2$, we will only have $\Phi = 0$ when the domain walls from $\phi_1$ and $\phi_2$ intersect. Hence, for $N = 2$, we expect strings to be ubiquitous. For $N = 3$, $\Phi = 0$ requires the domain walls from the three fields to intersect, which leads to the formation of monopoles.

For $N = 4$, $\Phi = 0$ requires the domain walls from the four fields to intersect, which while possible, is no longer topologically guaranteed. An alternative way of thinking about it is that the first two fields form strings, as do the second two fields. These strings can intersect, but do so with probability zero. However, there is nonzero probability density for two strings to approach within an arbitrarily close distance of each other. As $N$ increases, the probability density of having $N$ surfaces almost intersect decreases drastically, and smaller values of $\bar{\nu}$ becomes less and less likely.

For this paper, we assume that $\bar{\nu} > 0$ for all of our computations, and investigate the $\bar{\nu} \rightarrow 0$ limit only for $N \ge 4$, so as to avoid considerations of the densities associated with topological defects.

\section{Counting Stationary Points} \label{sec:nfields}

Consider a random field $\Phi$. Given a particular realization of $\Phi$, we can construct an expression to count the number of stationary points in $\Phi$ by utilizing the following formula from BBKS \cite{Bardeen1986}.\footnote{In the mathematical literature, this formula is known as the Kac-Rice formula \cite{Kac1943, Rice1945}, and was constructed some 40 years prior to BBKS, who appear to have discovered it independently.}
\begin{align}
N = \int_{{\cal R}^3} d^3 x \, \delta^3 (\partial_i \Phi) \mydet{\partial_i \partial_j \Phi} f(\lambda_i^\Phi)
\end{align}
The function $f(\lambda_i^\Phi)$ is a function of the eigenvalues of the Hessian matrix $\partial_i \partial_j \Phi$ and can be used to select the appropriate objects of interest, such as peaks or troughs. We restrict this formula to the number of stationary points with height $\nu = \sqrt{\Phi}$ by introducing a Dirac delta function.
\begin{align}
N = \int_{{\cal R}^3} d^3 x \int_0^\infty d\nu \, \delta^3 (\partial_i \Phi) \mydet{\partial_i \partial_j \Phi} f(\lambda_i^\Phi) \delta(\nu - \sqrt{\Phi})
\end{align}
Converting this to a number density of stationary points at a given height, we obtain
\begin{align} \label{eq:generaldensity}
\frac{d\N}{d\nu} = \delta^3 (\partial_i \Phi) \mydet{\partial_i \partial_j \Phi} f(\lambda_i^\Phi) \delta(\nu - \sqrt{\Phi}).
\end{align}

\subsection{Specializing to Chi-Squared Fields}

We now specialize to fields $\Phi$ constructed by
\begin{align} \label{eq:Phi}
\Phi = \sum_{\alpha = 1}^N \phi_\alpha^2 = \phi_\alpha \phi_\alpha
\end{align}
where $\phi_\alpha$ are $N$ centered independent Gaussian random fields with the same power spectrum (defined in Appendix \ref{app:stats}), and we have abbreviated the sum over the fields using Einstein summation convention. Note the following spatial derivatives of $\Phi$.
\begin{align}
\partial_i \Phi = 2 \phi_\alpha \partial_i \phi_\alpha,
\qquad 
\partial_i \partial_j \Phi = 2 \phi_\alpha \partial_i \partial_j \phi_\alpha + 2 \partial_i \phi_\alpha \partial_j \phi_\alpha
\end{align}

Combining Eq. \eqref{eq:generaldensity} with Eq. \eqref{eq:Phi}, we obtain
\begin{align}
\frac{d\N}{d\nu} &= \delta^3 (\phi_\alpha \partial_i \phi_\alpha) \mydet{\phi_\alpha \partial_i \partial_j \phi_\alpha + \partial_i \phi_\alpha \partial_j \phi_\alpha} f(\lambda_i^\Phi) \delta(\nu - \sqrt{\phi_\alpha \phi_\alpha}).
\end{align}

To compute the ensemble average for the number density, we need to integrate over all $\phi_\alpha$ configurations weighted by their probability density. To write out these integrals, we let $\eta_i^\alpha = \partial_i \phi_\alpha$ and $\zeta_{ij}^\alpha = \partial_i \partial_j \phi_\alpha$. Furthermore, we write $\vec{\phi} = (\phi_1, \ldots, \phi_n)$ as a vector in field space, as well as $\vec{\eta}_i$ and $\vec{\zeta}_{ij}$.
\begin{align} \label{eq:sofar}
\left<\frac{d\N}{d\nu}\right> &= 
\int d \vec{\phi} \, d \vec{\eta}_i \, d \vec{\zeta}_{ij} \,
\delta^3 (\vec{\phi} \cdot \vec{\eta}_i) 
\mydet{\vec{\phi} \cdot \vec{\zeta}_{ij} + \vec{\eta}_i \cdot \vec{\eta}_j} 
f(\lambda_i^\Phi) 
\delta\left(\nu - \sqrt{\vec{\phi} \cdot \vec{\phi}}\right) 
\prob(\phi_\alpha, \eta_i^\alpha, \zeta_{ij}^\alpha)
\end{align}
The function $\prob(\phi_\alpha, \eta_i^\alpha, \zeta_{ij}^\alpha)$ is the probability density associated with the fields at a point. It will be important later that it reflects the $O(N)$ rotation symmetry of $\Phi$. We need only integrate over the field values at a point, as the behavior of the field away from this point does not affect the integrand (and hence integrates out, leaving a factor of unity).

\subsection{Field Rotations}

To proceed, we exploit the fact that $\Phi$ has an $SO(N)$ rotation symmetry in field space (indeed, it has an $O(N)$ symmetry, but $SO(N)$ is sufficient for our present purpose). Under the action of such a rotation, the number density of peaks is unchanged. Let us multiply the integrand in Eq. \eqref{eq:sofar} by a factor of 1 in the form of Eq. \eqref{eq:haarresult}.
\begin{align}
\left<\frac{d\N}{d\nu}\right> &= 
\int d \vec{\phi} \, d \vec{\eta}_i \, d \vec{\zeta}_{ij} \,
\delta^3 (\vec{\phi} \cdot \vec{\eta}_i) 
\mydet{\vec{\phi} \cdot \vec{\zeta}_{ij} + \vec{\eta}_i \cdot \vec{\eta}_j}
f(\lambda_i^\Phi)
\delta\left(\nu - \sqrt{\vec{\phi} \cdot \vec{\phi}}\right) 
\prob(\phi_\alpha, \eta_i^\alpha, \zeta_{ij}^\alpha)
\nonumber \\
& \qquad \times \frac{|\vec{\phi}|^{N-1}}{\mathrm{Vol}[SO(N-1)]} \int_{SO(N)} (H^T dH) \, \delta([H^T \vec{\phi}]_1) \ldots \delta([H^T \vec{\phi}]_{N-1}) \Theta([H^T \vec{\phi}]_N)
\end{align}
$H$ is a $SO(N)$ rotation matrix, and $(H^T dH)$ is the Haar measure on $SO(N)$, described in Appendix \ref{app:haar}. Following the first example in Appendix \ref{app:examples}, we change the order of integration to integrate over the fields $\phi_\alpha$ before integrating over the rotation matrices. Next, we change our field integration variables by performing a rotation in field space by $H$, letting $\vec{\phi} = H \vec{\bar{\phi}}$, $\vec{\eta}_i = H \vec{\bar{\eta}}_i$ and $\vec{\zeta}_{ij} = H \vec{\bar{\zeta}}_{ij}$. Note that the Jacobian of these transformations is unity, and that dot products in field space are invariant under this transformation.
\begin{align}
\left<\frac{d\N}{d\nu}\right> &= 
\frac{1}{\mathrm{Vol}[SO(N-1)]} \int_{SO(n)} (H^T dH) \, \int d \vec{\bar{\eta}}_i \, d \vec{\bar{\zeta}}_{ij} \,
\nonumber \\
& \qquad \times \int d\vec{\bar{\phi}} \, |\vec{\bar{\phi}}|^{N-1} \delta(\bar{\phi}_1) \ldots \delta(\bar{\phi}_{N-1}) \Theta(\bar{\phi}_N)
\delta^3 (\vec{\bar{\phi}} \cdot \vec{\bar{\eta}}_i) 
\mydet{\vec{\bar{\phi}} \cdot \vec{\bar{\zeta}}_{ij} + \vec{\bar{\eta}}_i \cdot \vec{\bar{\eta}}_j}
\nonumber \\
& \qquad \qquad \times
f(\lambda_i^\Phi)
\delta\left(\nu - \sqrt{\vec{\bar{\phi}} \cdot \vec{\bar{\phi}}}\right) 
\prob(H \vec{\bar{\phi}}, H\vec{\bar{\eta}}_i, H \vec{\bar{\zeta}}_{ij})
\end{align}
The integrals over $d\bar{\phi}_1$ through $d\bar{\phi}_{N-1}$ are now trivial to perform by integrating over the delta functions.
\begin{align}
\left<\frac{d\N}{d\nu}\right> &= 
\frac{1}{\mathrm{Vol}[SO(N-1)]} \int_{SO(N)} (H^T dH) \, \int d \vec{\bar{\eta}}_i \, d \vec{\bar{\zeta}}_{ij} \, d \bar{\phi}_N \, |\bar{\phi}_N|^{N-1} \Theta(\bar{\phi}_N)
\delta^3 (\bar{\phi}_N \bar{\eta}^N_i) 
\nonumber \\
& \qquad \qquad \times
\mydet{\bar{\phi}_N \bar{\zeta}^N_{ij} + \vec{\bar{\eta}}_i \cdot \vec{\bar{\eta}}_j}
f(\lambda_i^\Phi)
\delta\left(\nu - \bar{\phi}_N\right) 
\prob(H \vec{\bar{\phi}}, H\vec{\bar{\eta}}_i, H \vec{\bar{\zeta}}_{ij})
\end{align}
Note that the Heaviside function cleans up the delta function $\delta(\nu - \bar{\phi}_N)$ by removing the absolute value on $\bar{\phi}_N$. We can now integrate over $\bar{\phi}_N$ by using this delta function.
\begin{align}
\left<\frac{d\N}{d\nu}\right> &= 
\frac{1}{\mathrm{Vol}[SO(N-1)]} \int_{SO(N)} (H^T dH) \, \int d \vec{\bar{\eta}}_i \, d \vec{\bar{\zeta}}_{ij} \, \nu^{N-1}
\delta^3 (\nu \bar{\eta}^N_i) 
\nonumber \\
& \qquad \qquad \times
\mydet{\nu \bar{\zeta}^N_{ij} + \vec{\bar{\eta}}_i \cdot \vec{\bar{\eta}}_j}
f(\lambda_i^\Phi)
\prob(H \vec{\bar{\phi}}, H\vec{\bar{\eta}}_i, H \vec{\bar{\zeta}}_{ij})
\end{align}
At this stage, we invoke the invariance of the probability density $P$ under a field rotation. With all of the fields rotating the same way, the probability density is unchanged, and so we can replace $H \vec{\bar{\phi}} \rightarrow \vec{\bar{\phi}}$, etc. Doing this removes the last dependence on $H$ in the integrand, which allows us to integrate over the $SO(N)$ group.
\begin{align}
\left<\frac{d\N}{d\nu}\right> &= 
\frac{\mathrm{Vol}[SO(N)]}{\mathrm{Vol}[SO(N-1)]} \nu^{N-1} \int d \vec{\bar{\eta}}_i \, d \vec{\bar{\zeta}}_{ij} \, 
\delta^3 (\nu \bar{\eta}^N_i) 
\mydet{\nu \bar{\zeta}^N_{ij} + \vec{\bar{\eta}}_i \cdot \vec{\bar{\eta}}_j}
f(\lambda_i^\Phi)
\prob(\vec{\bar{\phi}}, \vec{\bar{\eta}}_i, \vec{\bar{\zeta}}_{ij})
\end{align}
The ratio of volumes can be computed from Eqs. \eqref{eq:onvol} and \eqref{eq:sonvol}. We can now drop the bars on $\bar{\phi}$, $\bar{\eta}$ and $\bar{\zeta}$.
\begin{align}
\left<\frac{d\N}{d\nu}\right> &= \frac{2 \pi^{N/2}}{\Gamma\left(\frac{N}{2}\right)}
 \nu^{N-1} \int d \vec{\eta}_i \, d \vec{\zeta}_{ij} \, 
\delta^3 (\nu \eta^N_i) 
\mydet{\nu \zeta^N_{ij} + \vec{\eta}_i \cdot \vec{\eta}_j}
f(\lambda_i^\Phi)
\prob(\vec{\phi}, \vec{\eta}_i, \vec{\zeta}_{ij})
\end{align}
Finally, writing $\delta^3(\nu \eta_i^N) = \delta^3(\eta_i^N) / \nu^3$, we can integrate over $\eta_i^N$. Furthermore, we can integrate over each $\zeta_{ij}^\alpha$ for $\alpha < N$, which just yields another factor of unity from the probability distribution. The result is
\begin{align}
\left<\frac{d\N}{d\nu}\right> &= \frac{2 \pi^{N/2}}{\Gamma\left(\frac{N}{2}\right)}
\nu^{N-4} \int d \zeta^N_{ij} \, \prod_{\alpha = 1}^{N-1} (d \eta^\alpha_i) \, 
\mydet{\nu \zeta^N_{ij} + \sum_{\alpha = 1}^{N-1} \eta_i^\alpha \eta_j^\alpha}
f(\lambda_i^\Phi)
\prob(\vec{\phi}, \vec{\eta}_i, \zeta^N_{ij}).
\end{align}

\subsection{Probability Distribution}

Our probability density describes the probabilities for the $\phi_\alpha$ fields and their first and second derivatives. Having integrated over all $\zeta_{ij}^\alpha$ for $\alpha < N$, as well as some delta functions fixing various values of $\phi_\alpha$ and $\eta^\alpha_i$, we desire the joint probability density for $\vec{\phi} = (0, \ldots, 0, \nu)$, $\vec{\eta}_i$ with $\eta^N_i = 0$, and $\zeta^N_{ij}$. Using Eq. \eqref{eq:probcalc}, this becomes the following.
\begin{align}
\prob(\vec{\phi}, \vec{\eta}_i, \zeta_{ij})
&=
\prob(\phi_1 = 0) \ldots \prob(\phi_{N-1} = 0)
\left(\prod_{\alpha = 1}^{N-1}\left[\prob(\eta_1^\alpha)\prob(\eta_2^\alpha)\prob(\eta_3^\alpha)\right]\right) \prob(\eta_1^N = 0) \prob(\eta_2^N = 0) \prob(\eta_3^N = 0)
\nonumber \\
& \qquad \times \prob(\phi_N = \nu, \zeta_{11}, \zeta_{22}, \zeta_{33}) \prod_{i < j} \prob(\zeta_{ij})
\end{align}
Breaking this up into pieces and using Eqs. \eqref{eq:probabilities}, \eqref{eq:probcomplicated} and \eqref{eq:probQ}, we obtain
\begin{align}
\prob(\phi_1 = 0) \ldots \prob(\phi_{N-1} = 0)
&=
\frac{1}{(2 \pi)^{(N-1)/2} \sigma_0^{N-1}}
\\
\left(\prod_{\alpha = 1}^{N-1}\left[\prob(\eta_1^\alpha)\prob(\eta_2^\alpha)\prob(\eta_3^\alpha)\right]\right) \prob(\eta_1^N = 0) \prob(\eta_2^N = 0) \prob(\eta_3^N = 0)
&=
\left(\frac{3}{2 \pi}\right)^{3N/2} \frac{1}{\sigma_1^{3N}} \exp\left[- \frac{3}{2 \sigma_1^2} 
\sum_{i} \sum_{\alpha = 1}^{N-1} \eta_i^\alpha \eta_i^\alpha
\right]
\\
\prob(\phi_N = \nu, \zeta_{11}, \zeta_{22}, \zeta_{33}) \prod_{i < j} \prob(\zeta_{ij})
&=
\frac{1}{(2 \pi)^{7/2}} \frac{5^2 \times 3^3}{2 \sigma_2^5} \sqrt{\frac{5}{\sigma_2^2 \sigma_0^2 - \sigma_1^4}}
e^{-B/2}
\end{align}
with
\begin{align}
B &= 
\frac{\nu^2}{\sigma_0^2}
+ \frac{1}{\sigma_0^2 \sigma_2^2 - \sigma_1^4} \frac{\sigma_1^4}{\sigma_0^2} \left(
\nu
+ \frac{\sigma_0^2}{\sigma_1^2} \sum_i \zeta_{ii}
\right)^2
+ \frac{5}{2 \sigma_2^2} \left[3 \sum_{i, j} (\zeta_{ij})^2
- \left(\sum_{i} \zeta_{ii}\right)^2 \right] .
\end{align}

Before combining these results, let us define
\begin{align}
\gamma \equiv \frac{\sigma_1^2}{\sigma_0 \sigma_2}
\end{align}
in order to clean up the expressions. Note that $0 < \gamma < 1$ due to the inequality in Eq. \eqref{eq:gammaineq}. We also take the opportunity to write $\nu = \sigma_0 \bar{\nu}$.

The full probability density function is
\begin{align}
\prob(\vec{\phi}, \vec{\eta}_i, \zeta_{ij})
&=
\frac{1}{(2 \pi)^{2N + 3}}
\frac{5^{5/2} 3^{3N/2+3}}{2 \sqrt{1 - \gamma^2}} \frac{1}{\sigma_0^N \sigma_1^{3N} \sigma_2^6} 
e^{- \bar{\nu}^2/2} e^{-Q/2}
\end{align}
with
\begin{align} \label{eq:Q}
Q = 
\frac{3}{\sigma_1^2} \sum_{i} \sum_{\alpha = 1}^{N-1} \eta_i^\alpha \eta_i^\alpha
+ \frac{1}{1 - \gamma^2} \left(
\gamma \bar{\nu}
+ \frac{1}{\sigma_2} \sum_i \zeta_{ii}
\right)^2
+ \frac{5}{2 \sigma_2^2} \left[3 \sum_{i, j} (\zeta_{ij})^2
- \left(\sum_{i} \zeta_{ii}\right)^2 \right].
\end{align}
The full integral is then
\begin{align} \label{eq:prematrix}
\left<\frac{d\N}{d\bar{\nu}}\right> &= 
\frac{dP}{d\bar{\nu}}
C_N
\int d \zeta^N_{ij} \, \prod_{\alpha = 1}^{N-1} (d \eta^\alpha_i) \, 
\mydet{\sigma_0 \bar{\nu} \zeta^N_{ij} + \sum_{\alpha = 1}^{N-1} \eta_i^\alpha \eta_j^\alpha}
f(\lambda_i^\Phi)
e^{-Q/2}
\end{align}
where we use Eq. \eqref{eq:dpdnu} to introduce $dP/d\bar{\nu}$, and we define
\begin{align}
C_N = \frac{1}{(2 \pi)^{(3N + 6)/2}}
\frac{5^{5/2} \cdot 3^{3N/2+3}}{2 \sqrt{1 - \gamma^2}} \frac{1}{\sigma_0^3 \sigma_1^{3N} \sigma_2^6 \bar{\nu}^3}
\end{align}
to absorb the coefficients.

\subsection{Matrix Variables}

Having integrated over all of the $\phi_\alpha$ and a number of $\eta_i^\alpha$ and $\zeta_{ij}^\alpha$ variables, the integral depends only on $\zeta_{ij}^N$ and a number of $\eta_i^\alpha$ variables. It is convenient to move the $\zeta$ and $\eta$ variables into a matrix notation before moving on.

A key feature of Eq. \eqref{eq:prematrix} is that the integrand only depends on $\eta_i^\alpha$ in the combination given by 
\begin{align}
\sum_{\alpha = 1}^{N-1} \eta_i^\alpha \eta_j^\alpha \equiv M_{ij}
\end{align}
which form a $3 \times 3$ matrix with spatial indices that we call $M$.

We can also view $\eta^\alpha_i$ as a matrix with $\alpha$ indexing rows and $i$ indexing columns. Correspondingly, let $A_{\alpha i}=\eta^\alpha_i$. $A$ is then a $(N-1)\times 3$ matrix with a copy of the gradient vector $\eta^\alpha_i$ from each field comprising each row. Now, $M = A^T A$, as
\begin{align}
A^T A = \sum_{\alpha=1}^{N-1} A_{\alpha i} A_{\alpha j} = \sum_{\alpha = 1}^{N-1} \eta_i^\alpha \eta_j^\alpha = M.
\end{align}
The outer product decomposition of $M$ tells us that $M$ is positive semi-definite.

Completing the transition to matrix notation, let us consider $\zeta^N_{ij}$ as elements of a $3\times 3$ matrix $Z$. For convenience, let $(dZ) = \prod_{i \le j} d\zeta^N_{ij}$ and $(dA) = \prod_i \prod_{\alpha = 1}^{N-1} d\eta^\alpha_i$ be our differential elements\footnote{Appendix \ref{app:haar} constructs measures for different types of matrix integrals in terms of an exterior product. This is important for performing some of our matrix integration, but we have hidden all of the details of these manipulations in the appendices. For understanding the body of the paper, simply note that $(dX)$ refers to the product of infinitesimals of all of the independent components of the matrix $X$.}. We now re-write Eqs. \eqref{eq:Q} and \eqref{eq:prematrix} in terms of these matrix variables as
\begin{align}
\left<\frac{d\N}{d\bar{\nu}}\right> &=
\frac{dP}{d\bar{\nu}}
C_N
\int (dZ) \, (dA) \, 
\mydet{\sigma_0 \bar{\nu} Z + M}
f(\lambda_i^\Phi)
e^{-Q/2}
\end{align}
where $Q$ is now given by
\begin{align}
Q = 
\frac{3}{\sigma_1^2} \tr[M]
+ \frac{1}{1 - \gamma^2} \left(
\gamma \bar{\nu}
+ \frac{1}{\sigma_2} \tr[Z]
\right)^2
+ \frac{5}{2 \sigma_2^2} \left[3 \tr[Z^2]
- \left(\tr[Z]\right)^2 \right].
\end{align}

In order to tidy up our integral, we now rescale our variables. Let $A = \sigma_1 \bar{A} / \sqrt{3}$, $Z = \sigma_2 \bar{Z}$ and $\bar{M} = \bar{A}^T \bar{A}$, so that $M = \sigma_1^2 \bar{M}/3$. The integral becomes
\begin{align}
\left<\frac{d\N}{d\bar{\nu}}\right> &=
\frac{dP}{d\bar{\nu}}
S_N
\int (d\bar{Z}) \, (d\bar{A}) \, 
\mydet{ \frac{3 \bar{\nu}}{\gamma} \bar{Z} + \bar{M}}
f(\lambda_i^\Phi)
e^{-Q/2}
\end{align}
where
\begin{align}
S_N = C_N 
\frac{\sigma_1^6}{3^3}
\sigma_2^6
\left(\frac{\sigma_1}{\sqrt{3}}\right)^{3(N-1)}
=
\frac{1}{(2 \pi)^{(3N + 6)/2}}
\frac{5^{5/2} \cdot 3^{3/2}}{2 \sqrt{1 - \gamma^2}} \frac{\sigma_1^3}{\sigma_0^3}
\frac{1}{\bar{\nu}^3}
\end{align}
captures all of the resulting coefficients after this scaling. The quantity $Q$ becomes
\begin{align}
Q = \tr[\bar{M}]
+ \frac{1}{1 - \gamma^2} \left(
\gamma \bar{\nu}
+ \tr[\bar{Z}]
\right)^2
+ \frac{5}{2}
\left[3 \tr[\bar{Z}^2] - \left(\tr[\bar{Z}]\right)^2 \right].
\end{align}
Moving forwards, we drop the bars on $A$, $M$ and $Z$.

\subsection{Roadblocks to an Analytic Integral}

If we are going to have a hope of analytically computing this integral, it appears that we need to get control of the eigenvalues $\lambda^\Phi_i$ so that our region of integration will be well-defined. We attempted to proceed by changing variables from $Z$ to $H$ by $Z = H - \gamma M / 3 \bar{\nu}$, so that the eigenvalues of $H$ and the eigenvalues of $\Phi$ have the same sign, which will satisfy any Heaviside step functions that appear in $f(\lambda^\Phi_i)$. This would also simplify the determinant to just the product of eigenvalues. Writing $H = R^T \Lambda^H R$ and integrating over the rotation group then leads to a reasonably straightforward-looking expression that integrates over the eigenvalues of $\Lambda^H$ and the matrix $A$.

Unfortunately, two issues complicate the proceedings. Firstly, we have $Z^2$ appearing in $Q$, which means that the change of variables introduces $M^2$, which contains the components of $A$ to the fourth power. The integrand is no longer a Gaussian in these variables, and integrating over more than one component of $A$ becomes intractable. Secondly, the Gaussians in $\lambda_i^H$ are not centered about $\lambda_i^H = 0$, while any Heaviside step functions restrict $\lambda_i^H$ to integrate from 0 to $\pm \infty$. Furthermore, the integrand contains the absolute value of the determinant. Together, these mean that integrating over an eigenvalue always leads to an error function, which contains the other eigenvalues in the argument. Further integration is then intractable.

We took this as a sign that our integration must be computed numerically. Given that we had pseudo-Gaussian integrals in $3N$ dimensions over infinite domains, it looked like quadrature methods would be challenging at best, and so we decided on Monte Carlo integration. After performing some tests, we found that the numerical integration worked best when $Q$ remained quadratic in all of our variables, and so we abandoned the idea of transforming variables from $Z$ to $H$.

\subsection{Spatial Rotations}\label{sec:spatial}

Before constructing our numerical integral, we can perform one last analytic simplification. Our measure $(dZ)$ is a product of six differentials, each of which is integrated from $-\infty$ to $\infty$. In Appendix \ref{app:matrixint}, we construct a method of integrating over a symmetric matrix $H$ by writing $H = R \Lambda R^T$, with $R$ an orthogonal matrix and $\Lambda$ the eigenvalues of $H$, and integrating over the resulting orthogonal group and the space of eigenvalues. We now apply this result to our integral to simplify the integral over $Z$.

Let us write $Z = R \Lambda^Z R^T$, where $\Lambda^Z = \mathrm{diag}(\lambda^Z_1, \lambda^Z_2, \lambda^Z_3)$ and $R^T R = R R^T = \mathbbm{1}$. Using Eq. \eqref{eq:matrixint}, we can write 
\begin{align}
\int (dZ) f(\Lambda^Z) = \frac{1}{2^3 \cdot 3!} \int_{O(3)} (R^T dR) \int (d\Lambda^Z) \, \Delta(\Lambda^Z) f(\Lambda^Z)
\end{align}
where $(d\Lambda^Z) = d\lambda^Z_1 d\lambda^Z_2 d\lambda^Z_3$ and
\begin{align}
\Delta(\Lambda^Z) = |\lambda^Z_3 - \lambda^Z_2| |\lambda^Z_3 - \lambda^Z_1| |\lambda^Z_2 - \lambda^Z_1|
\end{align}
displays the phenomenon of ``eigenvalue repulsion'' in random matrices.

Applying this to our integral, we obtain
\begin{align}
\left<\frac{d\N}{d\bar{\nu}}\right> &=
\frac{dP}{d\bar{\nu}}
\frac{S_N}{2^3 \cdot 3!} \int_{O(3)} (R^T dR) \int (d\Lambda^Z) \, (dA) \, \Delta(\Lambda^Z)
\mydet{\frac{3 \bar{\nu}}{\gamma} R \Lambda^Z R^T + M}
f(\lambda_i^\Phi)
e^{-Q/2}
\end{align}
with
\begin{align}\label{eq:qrot}
Q = \tr[M]
+ \frac{1}{1 - \gamma^2} \left(
\gamma \bar{\nu}
+ \tr[\Lambda^Z]
\right)^2
+ \frac{5}{2}
\left[3 \tr[(\Lambda^Z)^2] - \left(\tr[\Lambda^Z]\right)^2 \right]
\end{align}
where we have used the cyclic property of traces to eliminate $R$ from $Q$. We clean up the determinant by defining $A = \bar{A} R^T$ and $\bar{M} = \bar{A}^T \bar{A}$, so that $M = A^T A = R \bar{A}^T \bar{A} R^T = R \bar{M} R^T$. We then have
\begin{align}
\mydet{\frac{3 \bar{\nu}}{\gamma} R \Lambda^Z R^T + M}
=
\mydet{\frac{3 \bar{\nu}}{\gamma} R \Lambda^Z R^T + R \bar{M} R^T}
=
\mydet{\frac{3 \bar{\nu}}{\gamma} \Lambda^Z + \bar{M}}.
\end{align}
Note also that $\tr{M} = \tr{\bar{M}}$ and that the Jacobian of an orthogonal transformation is unity, so that our measure is unchanged. This procedure corresponds to performing a global spatial rotation.

The integral is now independent of $R$, and so we can integrate over $O(3)$, yielding
\begin{align}
\int_{O(3)} (R^T dR) = \mathrm{Vol}[O(3)] = 16 \pi^2.
\end{align}
After the dust settles, the integral takes the following form, where we drop the bars on $\bar{M}$ and $\bar{A}$.
\begin{align} \label{eq:finalalaytic}
\left<\frac{d\N}{d\bar{\nu}}\right> &=
2 \pi^2 \frac{dP}{d\bar{\nu}}
S_N \frac{1}{6} \int (d\Lambda^Z) \, (dA) \, \Delta(\Lambda^Z)
\mydet{\frac{3 \bar{\nu}}{\gamma} \Lambda^Z + M}
f(\lambda_i^\Phi)
e^{-Q/2}
\end{align}
The quantity $Q$ is unchanged from Eq. \eqref{eq:qrot}.

\subsection{Integrating over dA}

For $N \in \{1, 2, 3\}$, Eq. \eqref{eq:finalalaytic} is our final analytic form. For $N=1$, there are no integrals to perform over $(dA)$, and $M = 0$. For $N=2$, we have
\begin{align}
\int(dA) = \int d\eta_1 \, d\eta_2 \, d\eta_3, \qquad M_{ij} = \eta_i \eta_j, \qquad \tr[M] = \eta_1^2 + \eta_2^2 + \eta_3^2
\end{align}
while for $N = 3$, these become
\begin{align}
\int(dA) = \int d\eta_1 \, d\eta_2 \, d\eta_3 \, d\xi_1 \, d\xi_2 \, d\xi_3, 
\quad M_{ij} = \eta_i \eta_j + \xi_i \xi_j, 
\quad \tr[M] = \eta_1^2 + \eta_2^2 + \eta_3^2 + \xi_1^2 + \xi_2^2 + \xi_3^2
\end{align}
where all integrals are taken from $-\infty$ to $\infty$, and we have used $\vec{\eta}$ and $\vec{\xi}$ to refer to the different first derivative fields.

For $N \ge 4$, $A$ is a rank 3 matrix with probability one over the domain of integration, which affords us some extra tricks. Note that $A$ has $3(N-1)$ independent components, but the integrand depends only on $A^T A$, which has 6 independent components. For $N \ge 4$, we are able to integrate out the extra information in $A$.

In Appendix \ref{app:inta}, we show in Eq. \eqref{eq:intA} that we can write\footnote{Note that we have $n = N-1$, where $n$ is the dimension defined in the appendix.}
\begin{align}
\int (dA) \, f(M) = 
\frac{8 \pi^{3(N-1)/2}}{\Gamma_3\left((N-1)/2\right)} \int (dT) \, a^3 b^2 c \, (abc)^{N-5} f(M)
\end{align}
where $\Gamma_m$ is the multivariate Gamma function, defined as
\begin{align}
\Gamma_m(x) = \pi^{m(m-1)/4} \prod_{i=1}^m \Gamma\left[x - \frac{1}{2} (i-1)\right]
\end{align}
and integrating over $(dT)$ corresponds to
\begin{align}
\int (dT) = \int_0^\infty da \int_0^\infty db \int_0^\infty dc \int_{-\infty}^{\infty} dd \int_{-\infty}^{\infty} de \int_{-\infty}^{\infty} df.
\end{align}
The matrix $M$ becomes
\begin{align}
M = \mat{
	a^2 & ad & af \\
	ad & b^2 + d^2 & be + df \\
	af & be + df & c^2 + e^2 + f^2
}
\end{align}
which has a particularly simple trace. The integral in Eq. \eqref{eq:finalalaytic} becomes
\begin{align} \label{eq:finalint}
\left<\frac{d\N}{d\bar{\nu}}\right> &=
2 \pi^2 \frac{dP}{d\bar{\nu}}
S_N
\frac{8 \pi^{3(N-1)/2}}{\Gamma_3\left((N-1)/2\right)}
\frac{1}{6} \int (d\Lambda^Z) \, (dT) \,
\Delta(\Lambda^Z) \,
a^2 b (abc)^{N-4}
\mydet{\frac{3 \bar{\nu}}{\gamma} \Lambda^Z + M}
f(\lambda_i^\Phi)
e^{-Q/2}
\end{align}
with
\begin{align} \label{eq:finalQ}
Q = a^2 + b^2 + c^2 + d^2 + e^2 + f^2
+ \frac{1}{1 - \gamma^2} \left(
\gamma \bar{\nu}
+ \tr[\Lambda^Z]
\right)^2
+ \frac{5}{2}
\left[3 \tr[(\Lambda^Z)^2] - \left(\tr[\Lambda^Z]\right)^2 \right].
\end{align}
Thus, for any $N \ge 4$, we have reduced our integral to a nine-dimensional integral.

\section{Numerical Integration} \label{sec:numintegrate}

We have now pushed our integral as far as we are analytically able to, and further progress must be made numerically. In this section, we develop a Monte Carlo scheme to compute our integral. We focus on the case of $N \ge 4$, but most of the techniques we describe here apply equally well for smaller $N$ too.

For a function $g(\Lambda^Z, M)$, define a Monte Carlo expectation value by
\begin{align} \label{eq:montecarlo}
\langle g(\Lambda^Z, M) \rangle_{MC} = \frac{1}{V_N} \frac{1}{6} \int (d\Lambda^Z) \int (dT) \,
\Delta(\Lambda^Z) a^2 b (abc)^{N-4}
e^{-Q / 2} g(\Lambda^Z, M)
\end{align}
where $V_N$ is a normalization constant and $Q$ is given by Eq. \eqref{eq:finalQ}. This integral can be straightforwardly evaluated using a Metropolis Hastings Monte Carlo integration scheme. To fix our normalization, we demand that $\langle 1 \rangle_{MC} = 1$, which yields
\begin{align} \label{eq:normalize}
V_N = \frac{1}{6} \int (d\Lambda^Z) \int (dT) \,
a^2 b (abc)^{N-4}
e^{-Q / 2}.
\end{align}
We now compute this integral.

The integral over $(d\Lambda^Z)$ integrates over the infinite range for each $\lambda^Z_i$. If we instead choose to order the eigenvalues such that $\lambda^Z_1 \le \lambda^Z_2 \le \lambda^Z_3$, then we can instead write
\begin{align}
\frac{1}{6} \int (d\Lambda^Z) = \int_{-\infty}^{\infty} d\lambda^Z_3 \int_{-\infty}^{\lambda^Z_3} d\lambda^Z_2 \int_{-\infty}^{\lambda^Z_2} d\lambda^Z_1
\end{align}
where the factor of 6 is a symmetry factor to compensate for the number of different ways the eigenvalues may be ordered. This ordering has the additional benefit of simplifying $\Delta(\Lambda^Z)$, which becomes
\begin{align}
\Delta(\Lambda^Z) = (\lambda^Z_3 - \lambda^Z_2) (\lambda^Z_3 - \lambda^Z_1) (\lambda^Z_2 - \lambda^Z_1)
\end{align}
without any absolute values.

We now introduce a clever coordinate transformation, adapted from Ref. \cite{Mahoux2001}. Assuming that our eigenvalues are ordered as above, we define
\begin{subequations}
\begin{align}
X &= \frac{\gamma \bar{\nu} + \lambda_1^Z + \lambda_2^Z + \lambda_3^Z}{\sqrt{1 - \gamma^2}}
\qquad &
\lambda^Z_1 &= \frac{X \sqrt{1 - \gamma^2} - \gamma \bar{\nu}}{3} - \frac{2}{3 \sqrt{5}} r \sin \left(\theta + \frac{\pi}{6}\right)
\\
x &= r \cos(\theta) = \frac{\sqrt{5}}{2} (2 \lambda_3^Z - \lambda_2^Z - \lambda_1^Z)
\qquad &
\lambda^Z_2 &= \frac{X \sqrt{1 - \gamma^2} - \gamma \bar{\nu}}{3} + \frac{2}{3 \sqrt{5}} r \sin \left(\theta - \frac{\pi}{6}\right)
\\
y &= r \sin(\theta) = \frac{\sqrt{15}}{2} (\lambda_2^Z - \lambda_1^Z)
\qquad &
\lambda^Z_3 &= \frac{X \sqrt{1 - \gamma^2} - \gamma \bar{\nu}}{3} + \frac{2}{3 \sqrt{5}} r \cos(\theta).
\end{align}
\end{subequations}
The range of $X$ is $-\infty < X < \infty$, while $x$, $y$ and $r$ range from 0 to $\infty$ thanks to the ordering of the eigenvalues. For $\theta$, note that
\begin{align}
\Delta(\Lambda^Z) = (\lambda_3^Z - \lambda_2^Z)(\lambda_3^Z - \lambda_1^Z)(\lambda_2^Z - \lambda_1^Z)
= \frac{2}{15^{3/2}} (3x^2 y - y^3)
= \frac{2}{15^{3/2}} r^3 \sin(3\theta)
\ge 0.
\end{align}
As each factor is positive by construction, the product must also be positive. Combined with $x \ge 0$ and $y \ge 0$, this implies the further constraint $y \le \sqrt{3} x$, which is equivalent to $0 \le \theta \le \pi/3$. The combination of $0 \le r$ and $0 \le \theta \le \pi / 3$ is then equivalent to $\lambda_1^Z \le \lambda_2^Z \le \lambda_3^Z$. In these variables, we have
\begin{align}
\frac{1}{1 - \gamma^2} \left(
\gamma \bar{\nu}
+ \tr[\Lambda^Z]
\right)^2 = X^2
\qquad \text{and} \qquad
\frac{5}{2}
\left[3 \tr[(\Lambda^Z)^2] - \left(\tr[\Lambda^Z]\right)^2 \right] = r^2
\end{align}
while the Jacobian determinant is simply
\begin{align}
\left|\frac{\partial(\lambda_1^Z, \lambda_2^Z, \lambda_3^Z)}{\partial(X, r, \theta)}\right| = \frac{2\sqrt{3} \sqrt{1 - \gamma^2}}{45} r.
\end{align}

Using these coordinates, we can write Eq. \eqref{eq:normalize} as
\begin{align}
V_N = 
\frac{4 \sqrt{1 - \gamma^2}}{5^{5/2} \cdot 3^3} \int_{-\infty}^\infty dX 
\int_0^\infty dr
\int_0^{\pi/3} d\theta \,
\int (dT) \,
r^4 \sin(3\theta)
a^2 b (abc)^{N-4}
e^{-Q/2}
\end{align}
with
\begin{align}
Q = a^2 + b^2 + c^2 + d^2 + e^2 + f^2 + r^2 + X^2.
\end{align}
Mathematica will straightforwardly compute this integral, yielding
\begin{align}
V_N =
\frac{2^{3(N-1)/2}}{5^{5/2} \cdot 3^3} \pi \sqrt{1 - \gamma^2} \Gamma_3\left(\frac{N-1}{2}\right).
\end{align}
While these variables are conducive to performing analytic integrals, we found that Monte Carlo integration in these variables converged significantly slower than in the $\lambda^Z_i$ variables.

Now that we are in possession of a normalized Monte Carlo expectation value, we can write the number density that we wish to compute in terms of it as
\begin{align} \label{eq:finalresult}
\left<\frac{d\N}{d\bar{\nu}}\right> &=
\alpha \frac{dP}{d\bar{\nu}}
\left\langle \mydet{\frac{3 \bar{\nu}}{\gamma} \Lambda^Z + M} f(\lambda_i^\Phi) \right\rangle_{MC}
\end{align}
where we define the surprisingly simple
\begin{align}
\alpha = 2 \pi^2 S_N \frac{8 \pi^{3(N-1)/2}}{\Gamma_3\left((N-1)/2\right)} V_N
= \frac{1}{(6 \pi)^{3/2}}
\frac{\sigma_1^3}{\sigma_0^3}
\frac{1}{\bar{\nu}^3}.
\end{align}
Note that our number density depends on a small number of parameters. It is directly proportional to $\sigma_1^3/\sigma_0^3$, which carries all of the units in the problem, and $\sigma_0$ alone in converting from $d{\cal N}/d\bar{\nu} \rightarrow d{\cal N}/d\nu$. Beyond these basic scaling factors, the expectation value depends on $N$, $\gamma$ and $\bar{\nu}$, but that is all.

\subsection{Analytic Integrals}

In programming a numerical integrator, it is useful to have something to test it against. Using similar techniques as for computing the normalization of the Monte Carlo integral, we can compute the following expectation values.
\begin{align}
\left\langle \text{min}(\lambda^Z_i) \right\rangle_{MC}
&=
- \frac{3}{\sqrt{10 \pi}} - \frac{\gamma \nu}{3}
\\
\left\langle \det \left(\frac{3 \bar{\nu}}{\gamma}\Lambda^Z\right) \right\rangle_{MC}
&=
3 \bar{\nu}^4 - \bar{\nu}^6
\\
\left\langle \det (M) \right\rangle_{MC}
&=
(N-1)(N-2)(N-3)
\end{align}

It turns out that we can analytically integrate a quantity very closely related to the number density of stationary points, which we call the \textit{signed number density}. This computes the number density of all stationary points, counting points with 1 or 3 positive eigenvalues positively, and 0 or 2 positive eigenvalues negatively, and is given by
\begin{align}
\left<\frac{d\N^\text{signed}}{d\bar{\nu}}\right> &=
\alpha \frac{dP}{d\bar{\nu}}
\left\langle \det \left(\frac{3 \bar{\nu}}{\gamma}\Lambda^Z + M\right) \right\rangle_{MC}.
\end{align}
Mathematica will compute this expectation value when written in terms of $X$, $r$ and $\theta$ as above. The result is
\begin{align}
\left\langle \det \left(\frac{3 \bar{\nu}}{\gamma}\Lambda^Z + M\right) \right\rangle_{MC}
&=
(N-1)(N-2)(N-3) - 3(N-1)^2 \bar{\nu}^2 + 3N \bar{\nu}^4 - \bar{\nu}^6
\\
\left<\frac{d\N^\text{signed}}{d\bar{\nu}}\right>
&=
\frac{\sigma_1^3}{\sigma_0^3} \frac{(N-1)(N-2)(N-3) - 3(N-1)^2 \bar{\nu}^2 + 3N \bar{\nu}^4 - \bar{\nu}^6}{2^{N/2 - 1} (6 \pi)^{3/2} \Gamma(N/2)}
\bar{\nu}^{N-4} e^{-\bar{\nu}^2/2}
.
\end{align}
Curiously, this is independent of $\gamma$. It turns out that this formula also holds for $N \in \{1, 2, 3\}$ by analytically continuing $N$, which we checked by explicit computation.

A topological theorem from Morse theory exists which states that the signed sum of all stationary points should be vanishing. It is unclear how the topological defects at low $N$ interact with this theorem, so we only expect it to hold for $N \ge 4$. Indeed, for all $N \ge 4$, we find that
\begin{align}
\int_0^\infty d\bar{\nu} \left<\frac{d\N^\text{signed}}{d\bar{\nu}}\right> = 0
\end{align}
satisfying the theorem.

\subsection{Maxima and Minima}

We finally turn to the factor of $f(\lambda^\Phi_i)$ that we have dragged through the entire computation. $\lambda^\Phi_i$ refers to the eigenvalues of the Hessian of our original $\Phi$ field. In order to count various types of stationary points, we are only interested in the \textit{sign} of the eigenvalues, and not their magnitude. The eigenvalues of the Hessian of $\Phi$ are proportional to the eigenvalues of $3 \bar{\nu} \Lambda^Z / \gamma + M$ by a positive constant, and hence the signs of the eigenvalues will correspond between the two matrices.

Given a point in the domain of integration, we would like to use the function $f(\lambda^\Phi_i)$ to determine what type of stationary point it is. In order to separate these different types of stationary points, we will write $f(\lambda^\Phi_i) = \Theta(\pm\lambda^\Phi_1) \Theta (\pm\lambda^\Phi_2) \Theta(\pm\lambda^\Phi_3)$ with signs chosen appropriately. In order for the point to be a minimum, we want the eigenvalues to all be positive, while counting maxima in the field requires all eigenvalues to be negative. We thus see how the signed number density counts different types of stationary points differently by ``undoing'' the absolute value of the determinant.

Since we only require the signs and not the eigenvalues themselves, we can use Sylvester's criterion, which states that a matrix is positive definite if and only if its leading principle minors are positive. The three leading principle minors of $3 \bar{\nu} \Lambda^Z / \gamma + M$ are
\begin{align}
\frac{3 \bar{\nu}}{\gamma} \lambda^Z_1 + a^2, 
\qquad 
\det \mat{
	\frac{3 \bar{\nu}}{\gamma} \lambda^Z_1 + a^2 & ad \\
	ad & \frac{3 \bar{\nu}}{\gamma} \lambda^Z_2 + b^2 + d^2
},
\quad \text{and} \quad
\det\left(\frac{3 \bar{\nu}}{\gamma} \Lambda^Z + M\right).
\end{align}
If all three minors are positive, we have a minimum, while if the first and third are negative with the second positive, we have a maximum. Otherwise, stationary points with two negative eigenvalues will have $\det(3 \bar{\nu} \Lambda^Z / \gamma + M) > 0$, while stationary points with one negative eigenvalue will have negative determinant. This means that it is computationally inexpensive to compute the signs of the eigenvalues, and hence determine the type of stationary point.

\subsection{Limits}

It is useful to have an analytic handle on the behavior of the number density of different types of stationary points in various limits. Unfortunately, only the crudest limits are amenable to analytic calculations, but these do give insight into how some of the number densities behave.

We begin by looking at the limit of large $\bar{\nu}$. In this limit, the vast majority of stationary points are maxima, and we can approximate
\begin{align}
\left\langle \mydet{\frac{3 \bar{\nu}}{\gamma} \Lambda^Z + M} f(\lambda_i^\Phi) \right\rangle_{MC}
\simeq - \left(\frac{3 \bar{\nu}}{\gamma}\right)^3 \left\langle \det(\Lambda^Z) \right\rangle_{MC} \simeq \bar{\nu}^6.
\end{align}
The number density of maxima then decays as
\begin{align}
\left<\frac{d\N^\text{max}}{d\bar{\nu}}\right> &=
\alpha \frac{dP}{d\bar{\nu}} \bar{\nu}^6
\simeq
\frac{1}{(6 \pi)^{3/2}}
\frac{\sigma_1^3}{\sigma_0^3}
\frac{1}{2^{N/2 - 1} \Gamma(N/2)} \bar{\nu}^{N+2} e^{-\bar{\nu}^2/2}
\end{align}
in this limit. Note that this is independent of $\gamma$, as it must be to give the correct contribution to the signed number density.

In the limit of small $\bar{\nu}$, the vast majority of stationary points are minima. We can then approximate
\begin{align}
\left\langle \mydet{\frac{3 \bar{\nu}}{\gamma} \Lambda^Z + M} f(\lambda_i^\Phi) \right\rangle_{MC}
\simeq \left\langle \det(M) \right\rangle_{MC} = (N-3) (N-2) (N-1).
\end{align}
The number density of minima then grows as
\begin{align}
\left<\frac{d\N^\text{min}}{d\bar{\nu}}\right> &=
\alpha \frac{dP}{d\bar{\nu}} (N-3) (N-2) (N-1)
\simeq
\frac{1}{(6 \pi)^{3/2}}
\frac{\sigma_1^3}{\sigma_0^3}
\frac{(N-3) (N-2) (N-1)}{2^{N/2 - 1} \Gamma(N/2)} \bar{\nu}^{N-3} e^{-\bar{\nu}^2/2}
.
\end{align}
Again, this is independent of $\gamma$. It is interesting to note the behavior at $\bar{\nu} = 0$. For $N=4$, this is a constant. At $N=5$, it is zero, but has a non-zero first derivative. At $N=6$, the first derivative is zero, but non-zero second derivative, and so on. This behavior is seen in the numerics.

We can also look at how the integrals behave as $\gamma$ approaches its limits. In the limit of $\gamma \rightarrow 1$, where the power spectrum is tightly peaked at a single wavelength, the exponential $e^{-Q/2}$ contains a term that limits to a Dirac delta function, forcing $\tr[\Lambda^Z] = - \bar{\nu}$. While this allows one of the $\lambda^Z_i$ integrals to be evaluated, it unfortunately does not simplify the computation of any of the other integrals.

The limit $\gamma \rightarrow 0$ is more amenable to analytic computation. This limit requires the power spectrum to decay slowly at large $k$\footnote{$\gamma = 0$ requires $\sigma_2$ to be divergent while $\sigma_1$ and $\sigma_0$ are convergent. Even if $\sigma_2$ is divergent, $\gamma$ can still be well-defined by taking the limit as a cutoff approaches $\infty$.}, creating a lot of power at high frequencies. This leads to a large amount of jitter, and a correspondingly large number of peaks, which manifests itself as a $1 / \gamma^3$ divergence in the number density of all types of stationary points.

In the small $\gamma$ limit, our integral simplifies significantly.
\begin{align}
\left<\frac{d\N}{d\bar{\nu}}\right> &\simeq
\alpha \frac{dP}{d\bar{\nu}}
\left(\frac{3 \bar{\nu}}{\gamma}\right)^3
\frac{1}{V_N} \frac{1}{6} \int (d\Lambda^Z) \int (dT) \,
\Delta(\Lambda^Z) a^2 b (abc)^{N-4}
e^{-Q / 2} \mydet{\Lambda^Z} \Theta(\pm \lambda^Z_i)
\\
Q &= a^2 + b^2 + c^2 + d^2 + e^2 + f^2
+ \frac{3}{2}\left(
5 \tr[(\Lambda^Z)^2] 
- \left(\tr[\Lambda^Z]\right)^2\right)
\end{align}
As the matrix $M$ is negligible in the determinant, we can pick the type of stationary point we want by choosing just the signs of $\lambda^Z_i$. The integrals over $(dT)$ can then be performed immediately, as those variables decouple completely from the rest of the integral. This yields
\begin{align}
\left<\frac{d\N}{d\bar{\nu}}\right> &\simeq
\alpha \frac{dP}{d\bar{\nu}}
\left(\frac{3 \bar{\nu}}{\gamma}\right)^3
\frac{3^3 \cdot 5^{5/2}}{8 \pi}
\frac{1}{6} \int (d\Lambda^Z) \,
\Delta(\Lambda^Z)
\left|\lambda^Z_1 \lambda^Z_2 \lambda^Z_3\right| \Theta(\pm \lambda^Z_i)
\exp\left[-\frac{3}{4}\left(5 \tr[(\Lambda^Z)^2] - \left(\tr[\Lambda^Z]\right)^2\right)\right].
\end{align}
The integrand possesses a symmetry $\Lambda^Z \rightarrow - \Lambda^Z$, except for the Heaviside step functions, which flip the sign of all eigenvalues. This implies that maxima and minima have equal number densities, as do both types of saddle points.

Focusing on minima for the moment, the integral becomes
\begin{align}
\frac{1}{6} \int (d\Lambda^Z) \,&
\Delta(\Lambda^Z)
\left|\lambda^Z_1 \lambda^Z_2 \lambda^Z_3\right| \Theta(\pm \lambda^Z_i)
\exp\left[-\frac{3}{4}\left(5 \tr[(\Lambda^Z)^2] - \left(\tr[\Lambda^Z]\right)^2\right)\right]
\\
&= \frac{1}{6} \int_0^\infty d\lambda^Z_1 d\lambda^Z_2 d\lambda^Z_3 \,
\Delta(\Lambda^Z)
\lambda^Z_1 \lambda^Z_2 \lambda^Z_3
\exp\left[-\frac{3}{4}\left(5 \tr[(\Lambda^Z)^2] - \left(\tr[\Lambda^Z]\right)^2\right)\right]
\\
&= 
\int_0^\infty d\lambda^Z_3 \int_0^{\lambda^Z_3} d\lambda^Z_2 \int_0^{\lambda^Z_2} d\lambda^Z_1 \,
(\lambda^Z_3 - \lambda^Z_2) (\lambda^Z_3 - \lambda^Z_1) (\lambda^Z_2 - \lambda^Z_1)
\lambda^Z_1 \lambda^Z_2 \lambda^Z_3
\nonumber\\
& \qquad \qquad \qquad \qquad
\times \exp\left[-\frac{3}{4}\left(5 \tr[(\Lambda^Z)^2] - \left(\tr[\Lambda^Z]\right)^2\right)\right]
\\
&= \frac{2}{3^6 \cdot 5^4} (29 \sqrt{2 \pi} - 12 \sqrt{3 \pi})
\end{align}
where we used Mathematica to carry out the final integration. This means that in the limit of small $\gamma$, the number density of minima and maxima is
\begin{align}
\left<\frac{d\N}{d\bar{\nu}}\right> &\simeq
\frac{1}{\gamma^3}
\frac{1}{(6 \pi)^{3/2}}
\frac{29 \sqrt{2} - 12 \sqrt{3}}{4 \cdot 5^{3/2} \sqrt{\pi}}
\frac{\sigma_1^3}{\sigma_0^3}
\frac{dP}{d\bar{\nu}}.
\end{align}
Note that all of the $N$ and $\bar{\nu}$ dependence comes from $dP/d\bar{\nu}$ in this expression.

We can also evaluate the result for saddle points. We work with two positive and one negative eigenvalue.
\begin{align}
\frac{1}{6} \int (d\Lambda^Z) \,&
\Delta(\Lambda^Z)
\left|\lambda^Z_1 \lambda^Z_2 \lambda^Z_3\right| \Theta(\pm \lambda^Z_i)
\exp\left[-\frac{3}{4}\left(5 \tr[(\Lambda^Z)^2] - \left(\tr[\Lambda^Z]\right)^2\right)\right]
\\
&= 
- \frac{3}{6} \int_0^\infty d\lambda^Z_1 \int_0^\infty d\lambda^Z_2 \int_{-\infty}^0 d\lambda^Z_3 \,
|\lambda^Z_1 - \lambda^Z_2| (\lambda^Z_1 - \lambda^Z_3) (\lambda^Z_2 - \lambda^Z_3)
\lambda^Z_1 \lambda^Z_2 \lambda^Z_3
\nonumber\\
& \qquad \qquad \qquad \qquad
\exp\left[-\frac{3}{4}\left(5 \tr[(\Lambda^Z)^2] - \left(\tr[\Lambda^Z]\right)^2\right)\right]
\\
&= \int_0^\infty d\lambda^Z_1 \int_0^{\lambda^Z_1} d\lambda^Z_2 \int_0^\infty d\lambda^Z_3 \,
(\lambda^Z_1 - \lambda^Z_2) (\lambda^Z_1 + \lambda^Z_3) (\lambda^Z_2 + \lambda^Z_3)
\lambda^Z_1 \lambda^Z_2 \lambda^Z_3
\nonumber\\
& \qquad \qquad \qquad \qquad
\exp\left[-\frac{3}{4}\left(5 \tr[(\Lambda^Z)^2] - \left(\lambda^Z_1 + \lambda^Z_2 - \lambda^Z_3\right)^2\right)\right]
\\
&= \frac{2}{3^6 \cdot 5^4} (29 \sqrt{2 \pi} + 12 \sqrt{3 \pi})
\end{align}
In the second line, a factor of 3 is introduced as a symmetry factor from picking $\lambda^Z_3$ to be negative. In the third line, we let $\lambda^Z_3 \rightarrow -\lambda^Z_3$ and choose $\lambda^Z_1 > \lambda^Z_2$. Mathematica again evaluated the integral to obtain the final result. The number density for each type of saddle point is then given by
\begin{align}
\left<\frac{d\N}{d\bar{\nu}}\right> &\simeq
\frac{1}{\gamma^3}
\frac{1}{(6 \pi)^{3/2}}
\frac{29 \sqrt{2} + 12 \sqrt{3}}{4 \cdot 5^{3/2} \sqrt{\pi}}.
\frac{\sigma_1^3}{\sigma_0^3}
\frac{dP}{d\bar{\nu}}.
\end{align}
Again note that the only $N$ and $\bar{\nu}$ dependence comes from $dP/d\bar{\nu}$. This is very similar to the result for minima/maxima, but with a sum instead of a difference of radicals in the numerator, which increases their density by a factor of 3.055 compared to minima/maxima.

This factor of $\sim 3$ can be understood by noting that in the $\gamma \rightarrow 0$ limit, at a stationary point, there are eight ways to choose the signs of the eigenvalues. One of them gives rise to a minimum, one to a maximum, and three to each of the types of stationary points. When taking eigenvalue repulsion into account this result changes only ever so slightly. Our numerical results confirm the factor of 3.055.

\section{Numerical Results}\label{sec:numerics}

\begin{figure}[b]
	\centering
	\includegraphics[width=0.9\textwidth]{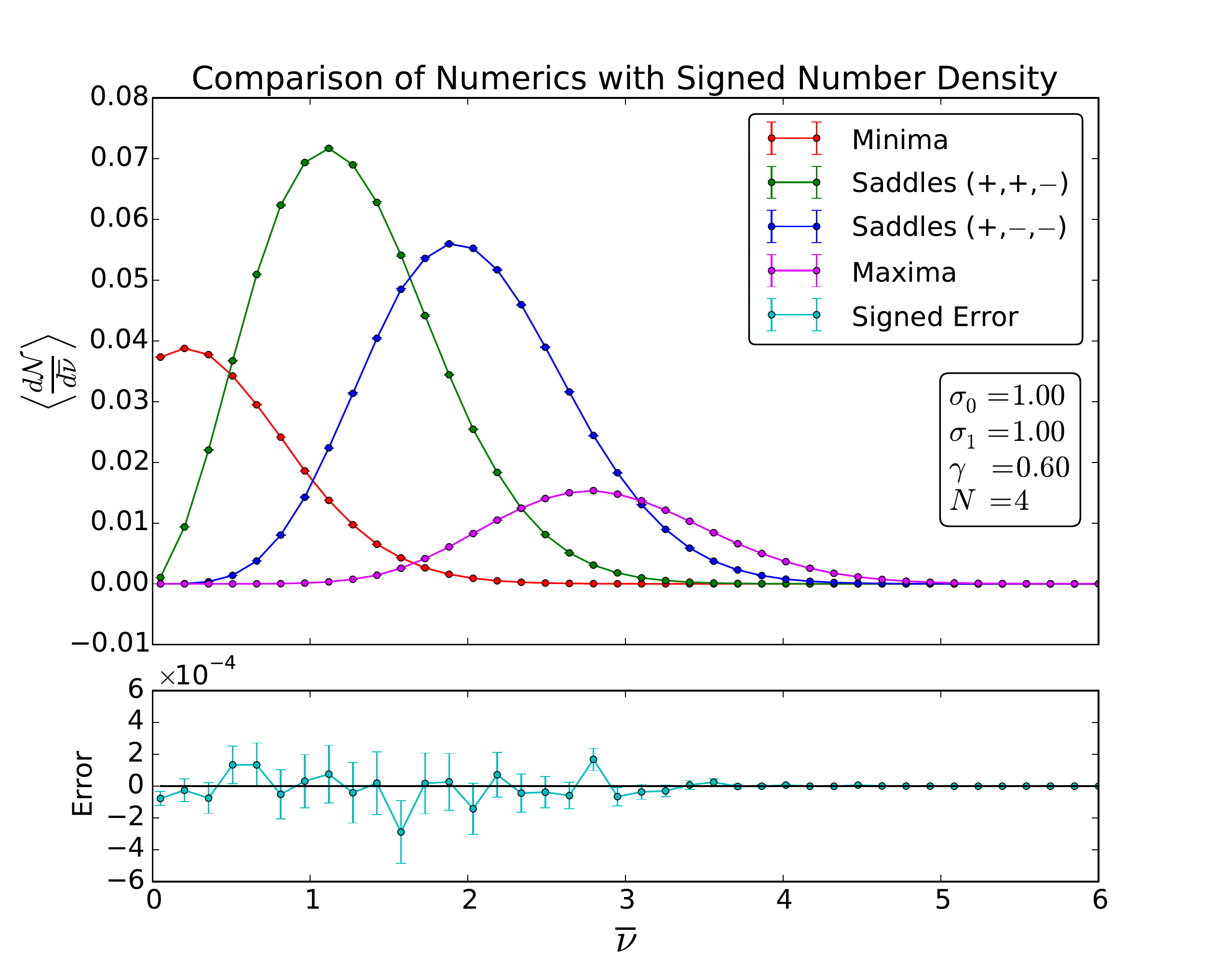}
	\caption{Plot of the number density of stationary points for each type of stationary point. The plot was made using our VEGAS implementation with $10^5$ samples. Statistical error bars are smaller than the resolution of the points in the main plot. In the lower plot, we display the error residual of the signed number density from combining all of the curves. Note that the error in the signed number density is greatest when cancellations occur between large numbers. The VEGAS implementation is very efficient: the data for this entire plot was computed in under 20 seconds on our laptops.}
	\label{fig:singleplot}
\end{figure}

\begin{figure}[p]
	\centering
	\includegraphics[width=\textwidth]{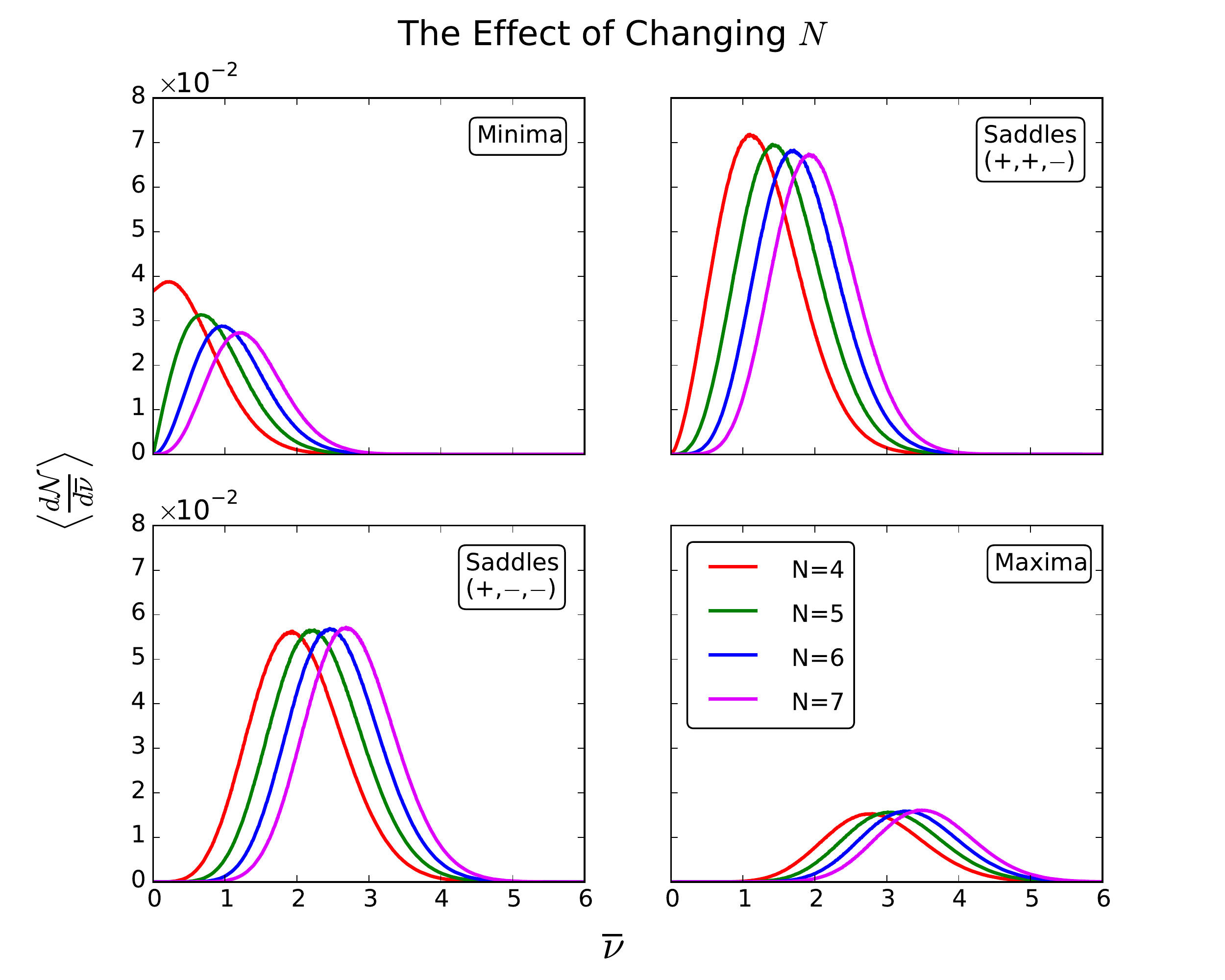}
	\caption{Plot of the number density of stationary points for different $N$. Each plot shows a different type of stationary point, with all plots using the same horizontal and vertical scales. These plots use $\sigma_0 = \sigma_1 = 1$ and $\gamma = 0.6$. Note the behavior of the minima at $\bar{\nu} = 0$, which is in excellent agreement with the predicted behavior.}
	\label{fig:Nplot}
\end{figure}

\begin{figure}[p]
	\centering
	\includegraphics[width=\textwidth]{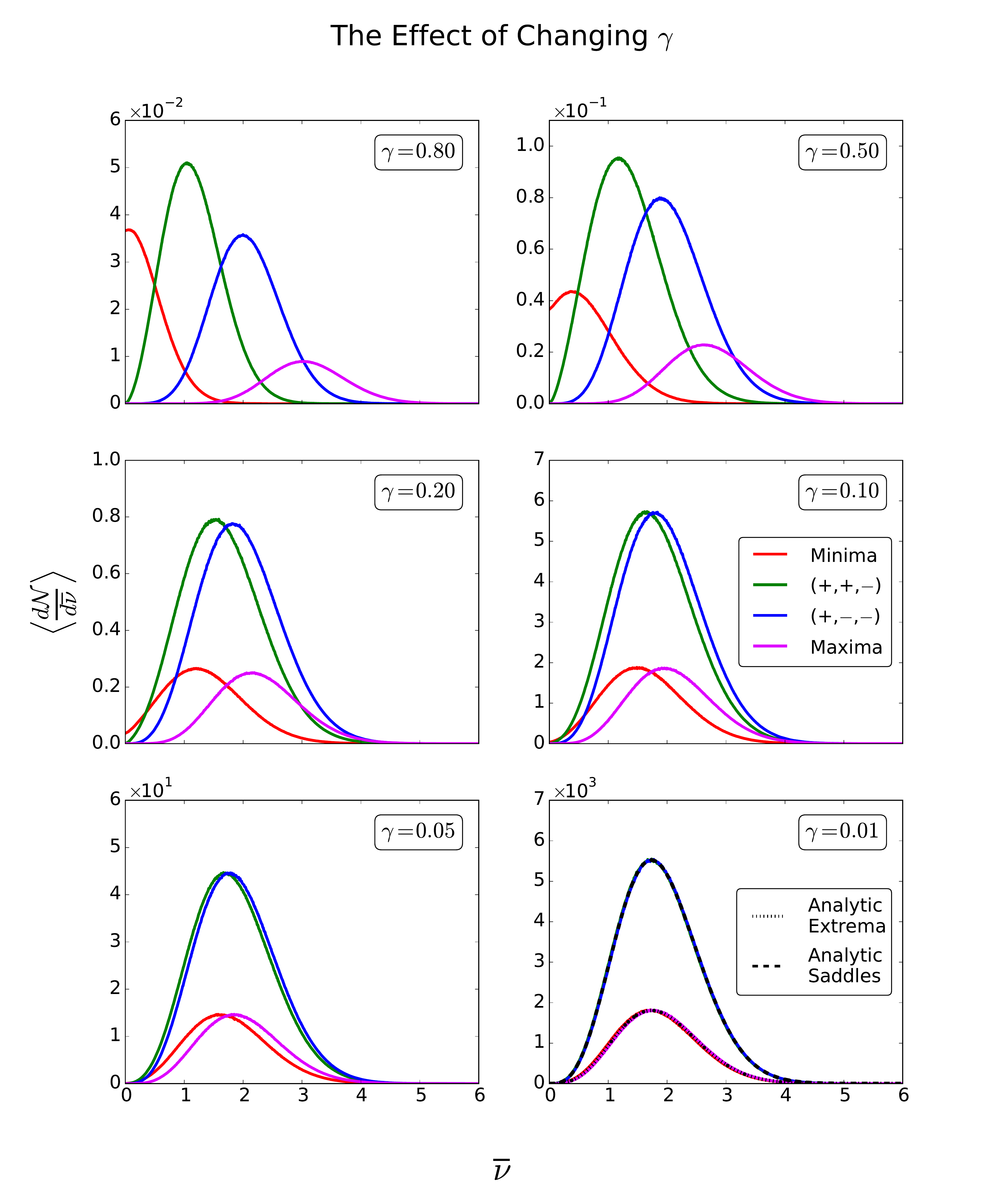}
	\caption{Plot of the number density of all stationary points at varying $\gamma$. These plots are computed with $N = 4$ and $\sigma_0 = \sigma_1 = 1$. Notice how the vertical scale of the plots changes rapidly with changing $\gamma$. As $\gamma$ decreases, the curves approach each other, until eventually they all peak at exactly the same location. In the $\gamma = 0.01$ figure, we also plot our analytic approximation for the curves, which is in excellent agreement with the numerical results. Also note that the $y$-intercept for the minima is independent of $\gamma$, in agreement with our analytic predictions.}
	\label{fig:gammaplot}
\end{figure}

We now have the tools to undertake the numerical computation of the number density of stationary points. To avoid topological defects, we restrict ourselves to $N \ge 4$. We avail ourselves of Eq. \eqref{eq:finalresult} to perform the computation and use $\sigma_1 = \sigma_2 = 1$ for the purposes of presenting results. Python code to perform the numerical integration has been made available online \cite{Bloomfield2016}.

We constructed a Metropolis-Hastings Monte Carlo integration code to compute the number density of each type of stationary point simultaneously. This allows us to compare to the analytical expression for the signed number density to get an idea for how accurate the results are. In regions of intermediate $\bar{\nu}$, where strong cancellations occur between the different types of stationary points, the error estimates for the signed number density tended to be somewhat large. This code was fairly slow, but had the benefit of being straightforward to implement.

We also implemented a VEGAS\footnote{The VEGAS algorithm is a highly optimized Monte Carlo integration algorithm that works very well for our particular integral. We made use of the excellent VEGAS python library.} version of the computation \cite{Lepage1978} which performed much better. This implementation had sub-percent convergence with only $10^5$ samples, and performed well throughout the range of $\bar{\nu}$. The results from both the MCMC and VEGAS implementations were equivalent, though the VEGAS implementation had superior convergence and speed.

Both implementations are set up to compute the number density of all types of stationary points simultaneously. The benefit to doing so is that the appropriate linear combination should agree with the signed number density, which gives an excellent check on the convergence. We demonstrate this convergence for a small sample size in Figure \ref{fig:singleplot}.

In Figure \ref{fig:Nplot}, we present the number density of each type of stationary point as $N$ changes from 4 to 7. Changing $N$ influences the curves primarily through the envelope of $dP/d\bar{\nu}$, which has the effect of shifting the curves to higher $\bar{\nu}$.

We demonstrate the effect of changing $\gamma$ in Figure \ref{fig:gammaplot}, where we plot the number density of minima for a variety of values of $\gamma$. Note that the number densities become larger and larger as $\gamma \rightarrow 0$, eventually diverging as $1/\gamma^3$. For $\gamma \lesssim 0.01$, our analytic estimate at $\gamma = 0$ is a very good approximation, as we show in the plot. The signed number density becomes difficult to numerically compute for small $\gamma$, as it arises as the difference of multiple large numbers that almost exactly cancel. Hence, error bars for the signed number density grow with decreasing $\gamma$, but the relative errors on the individual number densities remain small.

\section{Conclusions} \label{sec:conclusions}

We have presented a comprehensive investigation of the number density of stationary points in a field composed of the sum of squares of $N$ Gaussian random fields using generalizations of techniques used by BBKS for Gaussian fields. As $N < 4$ gives rise to topological defects, we have concentrated on the case for $N \ge 4$, but most of our formulas are applicable to the low $N$ case too.

Our final result is expressed as a 9-dimensional integral that must be evaluated numerically and computes the number density of stationary points at a given field height. Python source code has been provided to facilitate rapid and accurate numerical computations of the number density for each different type of stationary point, including both Metropolis-Hastings and VEGAS implementations.

The number density of stationary points depends on only 5 parameters. The parameter $N$ describes the number of fields under consideration, while $\bar{\nu}$ describes the dimensionless height of the field. The other three parameters $\sigma_0$, $\sigma_1$ and $\gamma$ are all related to moments of the power spectrum. Of these three parameters, only $\gamma$ appears in the integral, while $\sigma_0$ and $\sigma_1$ appear only as scaling parameters to define the appropriate scales in the result.

One expected feature of our results is that the number densities at a given height are directly proportional to the probability of a point having that height. This probability gives rise to the primary dependence of the result upon the number of fields $N$. At low heights, the integral scales roughly as $N^3$, while at high field heights, it is roughly independent of $N$.

The parameter $\gamma$ ranges between 0 and 1, where 0 is the limit of a slowly decaying tail in the power spectrum, and 1 is where all power is concentrated at a given wavelength. At $\gamma = 1$, the different types of stationary points are well-separated in $\bar{\nu}$, and are comparable in number density. As expected, low field heights have more minima, high field heights have more maxima, and different types of saddle points dominate in between. As $\gamma$ decreases, the peaks move towards each other, the number densities diverge, and saddle points become roughly three times more common than extrema. 

We were able to analytically evaluate various number densities in a number of limits, including $\gamma \rightarrow 0$, $\bar{\nu} \rightarrow 0$, and $\bar{\nu} \rightarrow \infty$. We were also able to analytically compute the signed number density. In all cases, these limits compared very well with our numerical results.

We envision using this result to investigate unusual events that result from extreme maxima or minima that follow this field structure in order to compute how often such events occur. An important caveat in doing so is what BBKS termed the ``cloud-in-cloud'' problem, where multiple peaks may give rise to only a single event. For example, in studying black hole collapse, large peaks may be required, but there are likely to be sub-peaks near the largest peak, especially if there is power at high frequencies. When a black hole forms, all nearby peaks collapse to form a single black hole, and care should be taken to not over-count the resulting number density of black holes. We are presently unaware of any solution to the cloud-in-cloud problem (in either the Gaussian case or the $\chi^2$ case), but hope that an appropriate filtering of the power spectrum will be able to resolve it.

\acknowledgments

We thank Mohammad Namjoo and Megan Russell for helpful conversations. This work is supported in part by the U.S. Department of Energy under grant Contract Number DE-SC0012567, and in part by MIT's Undergraduate Research Opportunities Program.

\bibliographystyle{utphys}
\bibliography{chisquared}

\appendix

\section{Gaussian Statistics} \label{app:stats}

In this appendix, we introduce the statistical description of our Gaussian fields. Our fields $\phi_\alpha(\vec{x})$ ($1 \le \alpha \le n$) are centered (zero mean) real Gaussian random fields with Fourier decomposition
\begin{align}
\phi_\alpha(\vec{x}) &= \int \frac{d^3k}{(\sqrt{2 \pi})^3} e^{i \vec{k} \cdot \vec{x}} \tilde{\phi}_\alpha (\vec{k})
= 
\int \frac{d^3k}{(\sqrt{2 \pi})^3} \frac{1}{2} \left[e^{i \vec{k} \cdot \vec{x}} \tilde{\phi}_\alpha (\vec{k}) + e^{-i \vec{k} \cdot \vec{x}} \tilde{\phi}_\alpha^\ast (\vec{k})\right] \,.
\end{align}
As $\phi_\alpha(\vec{x})$ is real, we have $\tilde{\phi}_\alpha^\ast(\vec{k}) = \tilde{\phi}_\alpha(-\vec{k})$. The field $\tilde{\phi}_\alpha(\vec{k})$ is a centered complex Gaussian random field, where the real and complex parts are uncorrelated and equidistributed, and the two point function of $\tilde{\phi}_\alpha(\vec{k})$ is given by
\begin{align}
\langle \tilde{\phi}_\alpha^\ast(\vec{k}) \tilde{\phi}_\beta(\vec{k}') \rangle &= \delta_{\alpha \beta}(2 \pi)^3 P(k) \delta^3(\vec{k} - \vec{k}')
\end{align}
which defines the stationary and isotropic power spectrum $P(k)$. Note that $\langle \phi(\vec{x}) \rangle = \langle \partial_i \phi(\vec{x}) \rangle = \langle \partial_i \partial_j \phi(\vec{x}) \rangle = 0$.

We compute the covariance $\langle \phi_\alpha(\vec{x}) \phi_\beta(\vec{y}) \rangle$ as
\begin{align}
\langle \phi_\alpha(\vec{x}) \phi_\beta(\vec{y}) \rangle
= \delta_{\alpha \beta}
\int d^3k \, e^{i \vec{k} \cdot (\vec{x} - \vec{y})} P(k).
\end{align}
We are mostly interested in the statistical properties of our fields in the coincidence limit. Because of the stationarity of the field, we can look at the point $\vec{x} = 0$ without loss of generality. In particular, we want to know the correlations between the field and its derivatives at a point. These correlations can be computed by evaluating the appropriate expectation values in Fourier space and using the definition of the power spectrum.
\begin{subequations}
\begin{align}
\langle \phi_\alpha(\vec{0}) \phi_\beta(\vec{0}) \rangle &= \delta_{\alpha \beta} 4 \pi \int_0^\infty dk \, k^2 P(k)
\\
\langle \phi_\alpha(\vec{0}) \partial_i \phi_\beta(\vec{0}) \rangle &= \langle \partial_i \phi_\alpha(\vec{0}) \partial_j \partial_m \phi_\beta(\vec{0}) \rangle = 0
\\
\langle \partial_i \phi_\alpha(\vec{0}) \partial_j \phi_\beta(\vec{0}) \rangle &= - \langle \phi_\alpha(\vec{0}) \partial_i \partial_j \phi_\beta(\vec{0}) \rangle = \delta_{\alpha \beta} \frac{4\pi}{3} \delta_{ij} \int_0^\infty dk \, k^4 P(k)
\\
\langle \partial_i \partial_j \phi_\alpha(\vec{0}) \partial_k \partial_l \phi_\beta(\vec{0}) \rangle &= \delta_{\alpha \beta} \frac{4\pi}{15} \left(
\delta_{ij} \delta_{kl}
+ \delta_{ik} \delta_{jl}
+ \delta_{il} \delta_{jk}
\right) \int_0^\infty dk \, k^6 P(k)
\end{align}
\end{subequations}
Note that we have used the following angular integrals to evaluate these covariances.
\begin{subequations}
\begin{align}
\int d\Omega \, n_i &= \int d\Omega \, n_i n_j n_k = 0
\\
\int d\Omega \, n_i n_j &= \frac{4\pi}{3} \delta_{ij}
\\
\int d\Omega \, n_i n_j n_k n_l &=
\frac{4\pi}{15} \left(
\delta_{ij} \delta_{kl}
+ \delta_{ik} \delta_{jl}
+ \delta_{il} \delta_{jk}
\right)
\end{align}
\end{subequations}

We define moments of the power spectrum by
\begin{align}
\sigma_n^2 = \int d^3k \, (k^2)^n P(k)
= 4 \pi \int_0^\infty dk \, (k^2)^{n+1} P(k).
\end{align}
Letting $\phi_\alpha(\vec{0}) \equiv \phi_\alpha$, $\partial_i \phi_\alpha(\vec{0}) \equiv \eta_i^\alpha$ and $\partial_i \partial_j \phi_\alpha(\vec{0}) \equiv \zeta_{ij}^\alpha$ as used in this paper, our correlations become
\begin{subequations}
\begin{align}
\langle \phi_\alpha \phi_\beta \rangle &= \delta_{\alpha \beta} \sigma_0^2
\\
\langle \phi_\alpha \eta_i^\beta \rangle &= \langle \eta_i^\alpha \zeta_{jk}^\beta \rangle = 0
\\
\langle \eta_i^\alpha \eta_j^\beta \rangle &= - \langle \phi_\alpha \zeta_{ij}^\beta \rangle = \frac{1}{3} \delta_{\alpha \beta} \delta_{ij} \sigma_1^2
\\
\langle \zeta_{ij}^\alpha \zeta_{kl}^\beta \rangle &= \frac{1}{15} \delta_{\alpha \beta} \left(\delta_{ij} \delta_{kl} + \delta_{ik} \delta_{jl} + \delta_{il} \delta_{jk}\right) \sigma_2^2.
\end{align}
\end{subequations}

Our Gaussian random variables are then $\phi_\alpha$, $\eta_i^\alpha$ and $\zeta_{ij}^\alpha$ for $1 \le \alpha \le N$ and spatial $i$ and $j$. We can see by inspection that each field has an autocorrelation. The only cross-correlations that arise are between $\phi_\alpha$ and $\zeta_{ii}^\alpha$, and between $\zeta_{ii}^\alpha$ and $\zeta_{jj}^\alpha$. Note that as these fields are multivariate Gaussian, they are completely described by these correlations.

\subsection{Probability Densities}

We can now construct the multivariate Gaussian probability distribution for all of our fields. The probability density for a centered Gaussian random variable is given by
\begin{align}
P(\vec{\phi}) = \frac{1}{|2 \pi \Sigma|^{1/2}} \exp \left[ - \frac{1}{2} \vec{\psi}^T \Sigma^{-1} \vec{\psi} \right]
\end{align}
where $\Sigma$ is the covariance matrix and $\vec{\psi}$ represents a vector of our random variables. The probability density factorizes for independent random variables, and so
\begin{align} \label{eq:probcalc}
P(\phi_\alpha, \eta_i^\alpha, \zeta_{ij}^\alpha) = \prod_{\alpha} \left[P(\phi_\alpha, \zeta_{11}^\alpha, \zeta_{22}^\alpha, \zeta_{33}^\alpha) \prod_{i \neq j} P(\zeta_{ij}^\alpha) \prod_i P(\eta_i^\alpha) \right].
\end{align}
We calculate the simpler factors first.
\begin{subequations} \label{eq:probabilities}
\begin{align}
P(\phi_\alpha) &= \frac{1}{\sqrt{2\pi} \sigma_0} \exp\left[- \frac{1}{2 \sigma_0^2} (\phi_\alpha)^2\right].
\\
P(\eta_i^\alpha) &= \sqrt{\frac{3}{2 \pi}} \frac{1}{\sigma_1} \exp\left[- \frac{3}{2 \sigma_1^2} (\eta_i^\alpha)^2\right]
\\
P(\zeta_{i\neq j}^\alpha) &= \sqrt{\frac{15}{2 \pi}} \frac{1}{\sigma_2} \exp\left[- \frac{15}{2 \sigma_2^2} (\zeta_{ij}^\alpha)^2\right]
\end{align}
\end{subequations}

The more complicated one to compute is $P(\phi, \zeta_{11}, \zeta_{22}, \zeta_{33})$, where we suppress the $\alpha$ index. Letting $\vec{\psi} = (\phi, \zeta_{11}, \zeta_{22}, \zeta_{33})$, the covariance matrix of $\vec{\psi}$ is given by
\begin{align} \label{eq:cov}
\Sigma = \mat{
\sigma_0^2 & - \frac{\sigma_1^2}{3} & - \frac{\sigma_1^2}{3} & - \frac{\sigma_1^2}{3} \\
- \frac{\sigma_1^2}{3} & \frac{\sigma_2^2}{5} & \frac{\sigma_2^2}{15} & \frac{\sigma_2^2}{15} \\
- \frac{\sigma_1^2}{3} & \frac{\sigma_2^2}{15} & \frac{\sigma_2^2}{5} & \frac{\sigma_2^2}{15} \\
- \frac{\sigma_1^2}{3} & \frac{\sigma_2^2}{15} & \frac{\sigma_2^2}{15} & \frac{\sigma_2^2}{5}
}.
\end{align}
The covariance matrix has determinant
\begin{align}
|\Sigma| = \frac{4}{675} \sigma_2^4 (\sigma_2^2 \sigma_0^2 - \sigma_1^4)
\end{align}
and inverse
\begin{align}
\Sigma^{-1} = \frac{1}{\sigma_0^2 \sigma_2^2 - \sigma_1^4} \mat{
\sigma_2^2 & \sigma_1^2 & \sigma_1^2 & \sigma_1^2 \\
\sigma_1^2 & \frac{6 \sigma_0^2 \sigma_2^2 - 5 \sigma_1^4}{\sigma_2^2} & \frac{5 \sigma_1^4 - 3 \sigma_0^2 \sigma_2^2}{2\sigma_2^2} & \frac{5 \sigma_1^4 - 3 \sigma_0^2 \sigma_2^2}{2\sigma_2^2} \\
\sigma_1^2 & \frac{5 \sigma_1^4 - 3 \sigma_0^2 \sigma_2^2}{2\sigma_2^2} & \frac{6 \sigma_0^2 \sigma_2^2 - 5 \sigma_1^4}{\sigma_2^2} & \frac{5 \sigma_1^4 - 3 \sigma_0^2 \sigma_2^2}{2\sigma_2^2} \\
\sigma_1^2 & \frac{5 \sigma_1^4 - 3 \sigma_0^2 \sigma_2^2}{2\sigma_2^2} & \frac{5 \sigma_1^4 - 3 \sigma_0^2 \sigma_2^2}{2\sigma_2^2} & \frac{6 \sigma_0^2 \sigma_2^2 - 5 \sigma_1^4}{\sigma_2^2}
}.
\end{align}
The covariance matrix should be positive definite. This requires the determinant to be positive definite, which in turn needs $\sigma_2 \sigma_0 - \sigma_1^2 > 0$. To see that this is true, consider
\begin{align} \label{eq:gammaineq}
0 &\le \frac{1}{2} \int d^3 k \, d^3 k' \, P(k) P(k') (k - k')^2 = \sigma_2 \sigma_0 - \sigma_1^2
\end{align}
where the initial inequality arises due to the integrand being positive, and is only saturated if $P(k) \equiv C \delta(k - k_0)$.

We therefore have
\begin{align}
P(\phi, \zeta_{11}, \zeta_{22}, \zeta_{33})
=
\frac{1}{(2 \pi)^2} \frac{15}{2 \sigma_2^2} \sqrt{\frac{3}{\sigma_2^2 \sigma_0^2 - \sigma_1^4}}
e^{-A / 2}
\end{align}
with $A = \vec{\psi}^T \Sigma^{-1} \vec{\psi}$, or
\begin{align}
A &= \frac{1}{\sigma_0^2 \sigma_2^2 - \sigma_1^4} \left[
\sigma_2^2 \phi^2
+ 2 \sigma_1^2 \phi \sum_i \zeta_{ii}
+ \frac{6\sigma_0^2 \sigma_2^2 - 5 \sigma_1^4}{\sigma_2^2} \sum_i (\zeta_{ii})^2 
+ \frac{5 \sigma_1^4 - 3 \sigma_0^2 \sigma_2^2}{\sigma_2^2} \sum_{i < j} \zeta_{ii} \zeta_{jj}
\right].
\end{align}
We combine this with the other components of $\zeta_{ij}$ to obtain
\begin{align} \label{eq:probcomplicated}
\prob(\phi, \zeta_{11}, \zeta_{22}, \zeta_{33}) \prob(\zeta_{12}) \prob(\zeta_{13}) \prob(\zeta_{23})
=
\frac{1}{(2 \pi)^{7/2}} \frac{5^2 \times 3^3}{2 \sigma_2^5} \sqrt{\frac{5}{\sigma_2^2 \sigma_0^2 - \sigma_1^4}}
e^{-B/2}
\end{align}
with
\begin{align}
B &= 
\frac{1}{\sigma_0^2 \sigma_2^2 - \sigma_1^4} \left[
\sigma_2^2 \phi^2
+ 2 \sigma_1^2 \phi \sum_i \zeta_{ii}
+ \frac{6\sigma_0^2 \sigma_2^2 - 5 \sigma_1^4}{\sigma_2^2} \sum_i (\zeta_{ii})^2 
+ \frac{5 \sigma_1^4 - 3 \sigma_0^2 \sigma_2^2}{\sigma_2^2} \sum_{i < j} \zeta_{ii} \zeta_{jj}
\right]
+ \frac{15}{\sigma_2^2} \sum_{i < j} (\zeta_{ij})^2.
\end{align}
This simplifies to
\begin{align}
B 
&= 
\frac{\phi^2}{\sigma_0^2}
+ \frac{1}{\sigma_0^2 \sigma_2^2 - \sigma_1^4} \frac{\sigma_1^4}{\sigma_0^2} \left(
\phi
+ \frac{\sigma_0^2}{\sigma_1^2} \sum_i \zeta_{ii}
\right)^2
+ \frac{5}{2 \sigma_2^2} \left[3 \sum_{i, j} (\zeta_{ij})^2
- \left(\sum_{i} \zeta_{ii}\right)^2 \right].
\label{eq:probQ}
\end{align}

\section{The Haar Measure} \label{app:haar}

In mathematics, the Haar measure provides a unique (up to multiplicative constant) measure on a group. In this appendix, we will perform an explicit construction of the Haar measure on the orthogonal group $O(n)$.

Before we begin, we should introduce some notation. Given a matrix $X$ of elements $(x_{ij})$ (note that we use (i, j) as (column, row)), define $dX$ to be the matrix of differential elements $(dx_{ij})$. If $X$ is an $n \times m$ matrix and $Y$ is an $m \times p$ matrix, then $d(XY) = X \cdot dY + dX \cdot Y$.

We will often need to take the exterior product of all of the independent differential elements in a matrix $dX$. This will be particularly useful when integrating over all of the elements of a matrix, as the integration measure is just the resulting differential form. For a general matrix of infinitesimals $dX$, define the bracket operation $(dX)$ by
\begin{align}
(dX) \equiv \bigwedge_{i,j} dx_{ij}.
\end{align}
If $X$ is a symmetric matrix, then we don't need to include all of the elements to construct the exterior product (which would vanish if we did). For symmetric matrices, let
\begin{align}
(dX) \equiv \bigwedge_{i \le j} dx_{ij}.
\end{align}
For antisymmetric matrices, we instead have
\begin{align}
(dX) \equiv \bigwedge_{i < j} dx_{ij}.
\end{align}
For matrices with zero entries such as diagonal or upper triangular matrices, the appropriate product ranges over the nonzero entries of the differential matrix.

Let $H$ be an $n \times n$ orthogonal matrix, such that $H^T H = \mathbbm{1}_{n \times n}$. We are interested in the matrix $H^T dH$. This is an antisymmetric matrix, as can be seen by differentiating the orthogonality condition to obtain $H^T dH = - dH^T H$. The columns of $H$ are $n$ orthonormal vectors, which we will denote $\{ \vec{h}_i \}$. This means we can write
\begin{align}
H = [ \vec{h}_1 \cdots \vec{h}_n], \qquad dH = [ d\vec{h}_1 \cdots d\vec{h}_n].
\end{align}
Hence, the elements of $H^T dH$ are given by
\begin{align}
[H^T dH]_{ij} = \vec{h}_j \cdot d\vec{h}_i.
\end{align}
Note that $\vec{h}_i \cdot d\vec{h}_i = 0$, as $\vec{h}_i \cdot \vec{h}_i = 1$ for the orthonormal columns. As a matrix, $H^T dH$ is the following.
\begin{align}
H^T dH = \mat{
	0 & - \vec{h}_2 \cdot d \vec{h}_1 & \ldots & - \vec{h}_n \cdot d\vec{h}_1 \\
	\vec{h}_2 \cdot d\vec{h}_1 & 0 & \ldots & - \vec{h}_n \cdot d\vec{h}_2 \\
	\vec{h}_3 \cdot d\vec{h}_1 & \vec{h}_3 \cdot d\vec{h}_2 & \ldots & - \vec{h}_n \cdot d\vec{h}_3 \\
	\vdots & \vdots & & \vdots \\
	\vec{h}_n \cdot d\vec{h}_1 & \vec{h}_n \cdot d\vec{h}_2 & \ldots & 0
}
\end{align}

We can take the bracket operation on $H^T dH$ to obtain the exterior product of its independent differential elements, remembering that $H^T dH$ is antisymmetric.
\begin{align}
(H^T dH) \equiv \bigwedge_{i < j} \vec{h}_j \cdot d\vec{h}_i
\end{align}
This quantity is invariant under left rotation $H \rightarrow QH$ by an element $Q \in O(n)$, as $H^T dH \rightarrow H^T Q^T Q dH = H^T dH$. To show that it is also invariant under right rotation, we need the following theorem:\footnote{This is proved in Chapter 2 of Ref. \cite{Muirhead2005}.}

\textit{\textbf{Theorem:} If $X = B^T Y B$ where $X$ and $Y$ are antisymmetric $n \times n$ matrices and $B$ is a non-singular $n \times n$ matrix, then}
\begin{align}
(X) = (\det B)^{n-1} (Y).
\end{align}

Consider $H$ under right rotation, $H \rightarrow HQ$. Then $H^T dH \rightarrow Q^T H^T dH Q$. Now, $(H^T dH) \rightarrow (Q^T H^T dH Q)$. Employing the above theorem, let $X = Q^T H^T dH Q$, such that $Y = H^T dH$ and $B = Q$. Then
\begin{align}
	(Q^T H^T dH Q) = (\det Q)^{n-1} (H^T dH).
\end{align}
But, $\det Q = \pm 1$ as $Q \in O(N)$, and hence $(H^T dH) \rightarrow \pm (H^T dH)$. The sign of the measure flips when performing an improper rotation that also includes a reflection ($\det Q = -1$), but this also changes the orientation of the coordinate system, and hence the sign is compensated by the sign of the orientation when we integrate over the differential form. Alternatively, we can think that the minus sign vanishes when taking the absolute value of the Jacobian of the transformation.

We now have a differential measure on $O(n)$ that is invariant under the left and right action of group members of $O(n)$. It is then meaningful to integrate this form over the group manifold to construct a measure on $O(n)$. Define the measure $\mu$ on $O(n)$ by
\begin{align}
	\mu({\cal D}) = \int_{\cal D} (H^T dH)
\end{align}
over the domain ${\cal D} \subset O(n)$. This is the Haar measure on $O(n)$. Note that because the differential form is invariant under the group action, so is the measure.
\begin{align}
	\mu(Q{\cal D}) = \mu({\cal D}Q) = \mu({\cal D}) \quad \forall \ Q \in O(n)
\end{align}

Before using the Haar measure, let us dissect its construction to develop some intuition about how it works. The quantity $(H^T dH)$ is the exterior product of the $n(n-1)/2$ components in the lower right triangle, and looks something like the following when written out completely.
\begin{align} \label{eq:haarwedges}
(H^T dH) = (\vec{h}_2 \cdot d\vec{h}_1) \wedge (\vec{h}_3 \cdot d\vec{h}_1) \wedge (\vec{h}_3 \cdot d\vec{h}_2) \wedge \ldots \wedge (\vec{h}_n \cdot d\vec{h}_{n-1})
\end{align}
Note that all factors are of the form $\vec{h}_j \cdot d\vec{h}_i$ with $i \neq j$. Each of these objects represent an independent combination of differentials from every other factor in the product. To see this, look for example at $\vec{h}_j \cdot d\vec{h}_1$. There are $n-1$ of these factors, each representing an orthonormal vector dotted into the differential. As each orthonormal vector is independent of the others, these $n-1$ orthonormal vectors form a basis for an $(n-1)$-dimensional vector space. The dot products are simply the independent components of $d\vec{h}_1$ in this vector space. The process repeats for each $d\vec{h}_i$. Hence, when we are integrating over the group, each component $\vec{h}_j \cdot d\vec{h}_i$ represents an independent differential to be integrating over. 

Note that an orthogonal matrix can be parametrized by $n(n-1)/2$ independent continuous parameters. The $n(n-1)/2$ independent differentials correspond to those parameters. In Eq. \eqref{eq:haarwedges}, it looks like there are no differentials $d\vec{h}_n$. However, as $\vec{h}_i \cdot \vec{h}_n = 0$ for $i \neq n$, we have $d \vec{h}_i \cdot \vec{h}_n = - \vec{h}_i \cdot d \vec{h}_n$, and so we actually are integrating over these differentials, just in disguise.

The quantity $\mu(O(n))$ represents the volume of the orthogonal group, and is given by\footnote{See Chapter 2 in Ref. \cite{Muirhead2005}.}
\begin{align} \label{eq:onvol}
\mathrm{Vol}[O(n)] &= \mu(O(n)) = \frac{2^n \pi^{n(n+1)/4}}{\prod_{j=1}^n \Gamma\left(\frac{j}{2}\right)}.
\end{align}
Integrating over a subgroup of $O(n)$ yields the volume of that subgroup. So, for example, integrating over $SO(n-1)$ rotations in $O(n)$ returns the volume of $SO(n-1)$.

The orthogonal group breaks down into two disjoint subsets $SO(n)$ and $SO^-(n)$, where $SO(n)$ is the subgroup of elements of $O(n)$ with determinant $+1$, and $SO^-(n)$ is the set of elements of $O(n)$ with determinant $-1$. Note that it is straightforward to construct a bijective map $\psi: SO(n) \rightarrow SO^-(n)$. As the union of $SO(n)$ and $SO^-(n)$ is $O(n)$, we must have
\begin{align}
\mu(O(n)) = \mu(SO(n)) + \mu(SO^-(n)).
\end{align}
Furthermore, due to the invariance of the Haar measure, we have $\mu(\psi(SO(n))) = \mu(SO(n))$, and so $\mu(SO(n)) = \mu(SO^-(n))$. Thus,
\begin{align} \label{eq:sonvol}
\mathrm{Vol}[SO(n)] = \frac{1}{2} \mathrm{Vol}[O(n)].
\end{align}

\subsection{Using the Haar measure}

Consider the vector $\vec{\phi} = (\phi_1, \ldots, \phi_n)$ with magnitude $|\phi| = \sqrt{\vec{\phi} \cdot \vec{\phi}}$. Now, consider the following integral.
\begin{align} \label{eq:haarintegral}
\int_{SO(n)} (H^T dH) \, \delta([H^T \vec{\phi}]_1) \ldots \delta([H^T \vec{\phi}]_{n-1}) \Theta([H^T \vec{\phi}]_n)
\end{align}
Noting that $[H^T \vec{\phi}]_i = \vec{h}_i \cdot \vec{\phi}$, these delta functions are restricting the integral to rotations $H$ that make $\vec{h}_i \cdot \vec{\phi} = 0$ for $1 \le i \le n-1$. We can decompose $\vec{\phi}$ into the orthonormal basis $\{\vec{h}_i\}$ as
\begin{align}
	\vec{\phi} = \sum_{i = 1}^n (\vec{\phi} \cdot \vec{h}_i) \vec{h}_i.
\end{align}
Thus, we are restricting the integral to rotations for which $\vec{h}_n = \pm \vec{\phi} / |\phi|$. The Heaviside step function then picks out the positive sign, so that $\vec{h}_n = \vec{\phi}/\phi$.

The $SO(n)$ group has $n(n-1)/2$ degrees of freedom; these delta functions fix $n-1$ of them. This makes sense, as the remaining $(n-1)(n-2)/2$ degrees of freedom are the degrees of freedom of the $SO(n-1)$ subgroup of rotations that leave $\vec{h}_n$ pointing in the same direction as $\vec{\phi}$. Note that modulo this $SO(n-1)$ subgroup, there is a unique rotation that fixes $\vec{h}_n = \vec{\phi} / |\phi|$, and so each of the delta functions in Eq. \eqref{eq:haarintegral} will have precisely one solution.

We would like to explicitly compute the integral in Eq. \eqref{eq:haarintegral}. Let us write the invariant form as
\begin{align}
	(H^T dH) = \quad {}& (\vec{h}_2 \cdot d\vec{h}_1)
	\nonumber \\
	{}\wedge {}& (\vec{h}_3 \cdot d\vec{h}_1) \wedge (\vec{h}_3 \cdot d\vec{h}_2)
	\nonumber\\
	&\qquad \vdots
	\nonumber\\
	{}\wedge {}& (\vec{h}_n \cdot d\vec{h}_1) \wedge \ldots \wedge (\vec{h}_n \cdot d\vec{h}_{n-1}).
\end{align}
Written in this suggestive manner, we see that the first line is the Haar measure on the $SO(2)$ subgroup, the first and second lines together are the Haar measure on the $SO(3)$ subgroup, etc. Taking all of the lines except the last, we obtain the Haar measure on the $SO(n-1)$ subgroup.

We can write our integral as an integral over the $SO(n-1)$ differential form, and a series of integrals over $(\vec{h}_n \cdot d\vec{h}_1) \wedge \ldots \wedge (\vec{h}_n \cdot d\vec{h}_{n-1})$ with delta functions (recall that each bracketed quantity is an independent quantity). Hence,
\begin{align}
	\int_{SO(n)} (H^T dH) & \delta([H^T \vec{\phi}]_1) \ldots \delta([H^T \vec{\phi}]_{n-1}) \Theta([H^T \vec{\phi}]_n)
\nonumber \\
	&=
	\left(\int_{SO(n-1)} (H^T dH)\right) \prod_{j=1}^{n-1} \left(\int (\vec{h}_n \cdot d\vec{h}_j) \delta(\vec{h}_j \cdot \vec{\phi})\right) \Theta([H^T \vec{\phi}]_n)
	\\
	&=
	\mathrm{Vol}[SO(n-1)] \prod_{j=1}^{n-1} \left(\int (\vec{h}_n \cdot d\vec{h}_j) \delta(\vec{h}_j \cdot \vec{\phi})\right) \Theta([H^T \vec{\phi}]_n)
\end{align}
where we use the technique of integrating over a differential form by ``dropping the wedges''.

We now look at the last few integrals. The effect of the delta and theta functions is to select $\vec{h}_n = \vec{\phi} / |\vec{\phi}|$. Thus, by the standard applications of delta functions, we are permitted to substitute this in for $\vec{h}_n$.
\begin{align}
\prod_{j=1}^{n-1} \left(\int (\vec{h}_n \cdot d\vec{h}_j) \delta(\vec{h}_j \cdot \vec{\phi})\right) \Theta(\vec{h}_n \cdot \vec{\phi})
= \frac{1}{|\vec{\phi}|^{n-1}}
\prod_{j=1}^{n-1} \left(\int (\vec{\phi} \cdot d\vec{h}_j) \delta(\vec{\phi} \cdot \vec{h}_j)\right)
\end{align}
We now substitute $u_j = \vec{\phi} \cdot \vec{h}_j$, such that $du_j = \vec{\phi} \cdot d\vec{h}_j$.
\begin{align}
\prod_{j=1}^{n-1} \left(\int (\vec{h}_n \cdot d\vec{h}_j) \delta(\vec{h}_j \cdot \vec{\phi})\right) \Theta(\vec{h}_n \cdot \vec{\phi})
&= \frac{1}{|\vec{\phi}|^{n-1}}
\prod_{j=1}^{n-1} \left(\int du_j \delta(u_j) \right)
\\
&= \frac{1}{|\vec{\phi}|^{n-1}}
\end{align}
Hence, the integral \eqref{eq:haarintegral} evaluates to
\begin{align}
\int_{SO(n)} (H^T dH) \, \delta([H^T \vec{\phi}]_1) \ldots \delta([H^T \vec{\phi}]_{n-1}) \Theta(\vec{h}_n \cdot \vec{\phi})
= \frac{\mathrm{Vol}[SO(n-1)]}{|\vec{\phi}|^{n-1}}.
\end{align}
We can thus write
\begin{align} \label{eq:haarresult}
1 = \frac{|\vec{\phi}|^{n-1}}{\mathrm{Vol}[SO(n-1)]} \int_{SO(n)} (H^T dH) \, \delta([H^T \vec{\phi}]_1) \ldots \delta([H^T \vec{\phi}]_{n-1}) \Theta([H^T \vec{\phi}]_n).
\end{align}

\subsection{Examples} \label{app:examples}

We now demonstrate two uses of the Haar measure. In the first, we consider an integral in $n$ dimensions of a function of radius only.
\begin{align}
A = \int dx_1 \ldots dx_n f(r)
\end{align}
We can multiply the integrand by one in the form of Eq. \eqref{eq:haarresult}, where we take our vector $\vec{\phi} = \vec{x} = (x_1, \ldots, x_n)$ with $|\vec{\phi}| = r$.
\begin{align}
A = \int dx_1 \ldots dx_n f(r) \frac{r^{n-1}}{\mathrm{Vol}[SO(n-1)]} \int_{SO(n)} (H^T dH) \, \delta([H^T \vec{x}]_1) \ldots \delta([H^T \vec{x}]_{n-1}) \Theta(\vec{h}_n \cdot \vec{x})
\end{align}
We now change the order of integration, and integrate over $dx_i$ first.
\begin{align}
A = \int_{SO(n)} (H^T dH) \, \int dx_1 \ldots dx_n f(r) \frac{r^{n-1}}{\mathrm{Vol}[SO(n-1)]} \delta([H^T \vec{x}]_1) \ldots \delta([H^T \vec{x}]_{n-1}) \Theta(\vec{h}_n \cdot \vec{x})
\end{align}
We perform a change of variables $\vec{x} = H \vec{y}$. As $H \in SO(n)$, the Jacobian for this transformation is unity, and $r = |\vec{y}|$. Note also that $\vec{h}_n \cdot H = (0, \ldots, 0, 1)$, so $\vec{h}_n \cdot H \cdot \vec{y} = y_n$.
\begin{align}
A &= \int_{SO(n)} (H^T dH) \, \int dy_1 \ldots dy_n f(r) \frac{r^{n-1}}{\mathrm{Vol}[SO(n-1)]} \delta(y_1) \ldots \delta(y_{n-1}) \Theta(y_n)
\\
&= \int_{SO(n)} (H^T dH) \, \int_0^\infty dy_n f(y_n) \frac{y_n^{n-1}}{\mathrm{Vol}[SO(n-1)]}
\end{align}
Now, there is no dependence on $H$ left, so we can perform that integral, which just yields the volume of $SO(n)$. We use Eqs. \eqref{eq:onvol} and \eqref{eq:sonvol} to compute the ratio of the volumes, and let $y_n \rightarrow r$.
\begin{align}
A = \frac{\mathrm{Vol}[SO(n)]}{\mathrm{Vol}[SO(n-1)]} \int_0^\infty dr \, r^{n-1} f(r) 
= \frac{2 \pi^{n/2}}{\Gamma\left(\frac{n}{2}\right)} \int_0^\infty dr \, r^{n-1} f(r) 
\end{align}
Note that the coefficient is simply the surface area of an $n$-dimensional sphere (with surface dimension $n-1$) with unit radius, as we would expect from writing $dx_1 \ldots dx_n = r^{n-1} dr d\Omega$ and performing the integration over $\Omega$. The benefit of integrating over the rotation group is that it provides a transparent way to rotate other fields by $H$, which is complicated when simply transforming to spherical polar coordinates.

The second example we present is an explicit example of constructing the Haar measure. Consider $SO(2)$. Matrices in $SO(2)$ can be parametrized by a single angle $\theta$ as
\begin{align}
H(\theta) = \mat{
\cos \theta & - \sin \theta \\
\sin \theta & \cos \theta	
} = \mat{\vec{h}_1 & \vec{h}_2}
\end{align}
with $0 \le \theta < 2 \pi$. The quantity $(H^T dH)$ is
\begin{align}
(H^T dH) 
= 
\vec{h}_2 \cdot d \vec{h}_1
=
\mat{- \sin \theta \\ \cos \theta} \cdot \mat{- \sin \theta d\theta \\ \cos \theta d\theta}
= d\theta.
\end{align}
Hence,
\begin{align}
\mathrm{Vol}[SO(2)] = \int (H^T dH) = \int_0^{2\pi} d\theta = 2 \pi
\end{align}
as expected.

\section{Integrating Over Symmetric Matrices} \label{app:matrixint}

In this appendix, we construct a method for integrating over the space of all symmetric $n \times n$ matrices by separately integrating over eigenvalues and a rotation group. The following makes extensive use of the Haar measure, which is constructed in Appendix \ref{app:haar}.

Consider a symmetric $n \times n$ matrix $H$ with elements $h_{ij}$, that is being integrated over as
\begin{align}
\int (dH) \, f(H) = \int dh_{11} dh_{12} dh_{13} dh_{22} dh_{23} dh_{33} \, f(H)
\end{align}
with each component integrated from $-\infty$ to $\infty$. The bracket notation $(dH)$ is described in Appendix \ref{app:haar}.

As $H$ is symmetric, it can be diagonalized using orthogonal matrices as $H = R \Lambda R^T$ where $R$ is orthogonal and $\Lambda$ is a diagonal matrix of the eigenvalues of $H$. We would like to write $(dH)$ in terms of parameters describing $R$ and $\Lambda$ and then integrate over those. However, we must be careful to compensate for any overcounting appropriately.

Consider a specific matrix $H$ which we decompose into $R$ and $\Lambda$. There are multiple choices for $R$ and $\Lambda$ that lead to the same $H$, and we need to determine what that overcounting factor is. There are $n$ eigenvalues to place into $\Lambda$, with $n!$ ways of doing so. Once $\Lambda$ is chosen, the ordering of the eigenvectors required to construct the appropriate $R$ matrix is fixed\footnote{This assumes that there are no degenerate eigenvalues. We are free to ignore the case of degenerate eigenvalues, as the space of degenerate eigenvalues is $\mathbb{R}^{n-1}$ (or lower), which has measure zero in $\mathbb{R}^n$.}, but we still have a choice of signs multiplying those eigenvectors. This leads to a $2^n$ degeneracy. Hence, if we integrate over all $R$ and $\Lambda$, the overcounting factor will be $2^n \cdot n!$.

We now construct the Jacobian of the transformation from $(dH)$ to $(dR)\,(d\Lambda)$. We begin with a theorem\footnote{This is derived in Chapter 2 of Ref. \cite{Muirhead2005}.}. 

\textit{\textbf{Theorem:} If $X = AYA^T$ for $X$ and $Y$ symmetric $n \times n$ matrices and $A$ a nonsingular $n \times n$ matrix, then}
\begin{align}
(dX) = \det(A)^{n+1} (dY).
\end{align}

Let $X = A H A^T$. Applying the theorem, we obtain
\begin{align}
(d(A H A^T)) = \det(A)^{n+1} (dH)
\end{align}
where we note that $d(A H A^T) = A dH A^T$. We now let $A = R^T$, so that
\begin{align}
(R^T dH R) = \det(R)^{n+1} (dH).
\end{align}
Note that $\det(R) = \pm 1$, so $(dH) = \pm (R^T dH R)$. We now write $H = R \Lambda R^T$, and expand its differential as
\begin{align}
dH = dR \Lambda R^T + R d\Lambda R^T + R \Lambda dR^T
\end{align}
so that
\begin{align}
(dH) = \pm (R^T dR \Lambda + d\Lambda + \Lambda dR^T R)
\end{align}
where the right hand side uses the bracket operation for constructing differential forms.

Let $\Lambda = \mathrm{diag}(\lambda_1, \ldots, \lambda_n)$ and $\vec{r}_i$ be the columns of $R$. It is straightforward to see that
\begin{align}
R^T dR \Lambda = \mat{
	0 & - \lambda_2 \vec{r}_2 \cdot d \vec{r}_1 & \ldots & - \lambda_n \vec{r}_n \cdot d\vec{r}_1 \\
	\lambda_1 \vec{r}_2 \cdot d\vec{r}_1 & 0 & \ldots & - \lambda_n \vec{r}_n \cdot d\vec{r}_2 \\
	\lambda_1 \vec{r}_3 \cdot d\vec{r}_1 & \lambda_2 \vec{r}_3 \cdot d\vec{r}_2 & \ldots & - \lambda_n \vec{r}_n \cdot d\vec{r}_3 \\
	\vdots & \vdots & & \vdots \\
	\lambda_1 \vec{r}_n \cdot d\vec{r}_1 & \lambda_2 \vec{r}_n \cdot d\vec{r}_2 & \ldots & 0
}
\end{align}
while
\begin{align}
\Lambda dR^T R = \mat{
	0 & \lambda_1 \vec{r}_2 \cdot d \vec{r}_1 & \ldots & \lambda_1 \vec{r}_n \cdot d\vec{r}_1 \\
	- \lambda_2 \vec{r}_2 \cdot d\vec{r}_1 & 0 & \ldots & \lambda_2 \vec{r}_n \cdot d\vec{r}_2 \\
	- \lambda_3 \vec{r}_3 \cdot d\vec{r}_1 & - \lambda_3 \vec{r}_3 \cdot d\vec{r}_2 & \ldots & \lambda_3 \vec{r}_n \cdot d\vec{r}_3 \\
	\vdots & \vdots & & \vdots \\
	- \lambda_n \vec{r}_n \cdot d\vec{r}_1 & - \lambda_n \vec{r}_n \cdot d\vec{r}_2 & \ldots & 0
}.
\end{align}
Hence, we have
\begin{align}
R^T dR \Lambda + d\Lambda + \Lambda dR^T R
= \mat{
	d \lambda_1 & (\lambda_1 - \lambda_2) \vec{r}_2 \cdot d \vec{r}_1 & \ldots & (\lambda_1 - \lambda_n) \vec{r}_n \cdot d\vec{r}_1 \\
	(\lambda_1 - \lambda_2) \vec{r}_2 \cdot d\vec{r}_1 & d \lambda_2 & \ldots & (\lambda_2 - \lambda_n) \vec{r}_n \cdot d\vec{r}_2 \\
	(\lambda_1 - \lambda_3) \vec{r}_3 \cdot d\vec{r}_1 & (\lambda_2 - \lambda_3) \vec{r}_3 \cdot d\vec{r}_2 & \ldots & (\lambda_3 - \lambda_n) \vec{r}_n \cdot d\vec{r}_3 \\
	\vdots & \vdots & & \vdots \\
	(\lambda_1 - \lambda_n) \vec{r}_n \cdot d\vec{r}_1 & (\lambda_2 - \lambda_n) \vec{r}_n \cdot d\vec{r}_2 & \ldots & d \lambda_n
}
\end{align}
which is symmetric, as expected. We can then construct $(dH)$ as
\begin{align}
(dH) = \pm \prod_{i < j} (\lambda_i - \lambda_j) (R^T dR) \prod_k d \lambda_k
\end{align}
where $(R^T dR)$ is the Haar measure. As we need the absolute value of the Jacobian for the change of variables, we can thus write
\begin{align} \label{eq:matrixint}
\int (dH) \, f(H) = \frac{1}{2^n \cdot n!} \int_{O(n)} (R^T dR) \int (d\Lambda) \, \Delta(\Lambda) f(R \Lambda R^T)
\end{align}
where each eigenvalue is integrated from $-\infty$ to $\infty$, the Haar measure is integrated over $O(n)$, and we define
\begin{align}
\Delta(\Lambda) = \prod_{i < j} | \lambda_i - \lambda_j |.
\end{align}
Note that we divide by the overcounting factor because we are integrating over all $R$ and $\Lambda$. This result is particularly useful if $f(H)$ depends only on the eigenvalues of $H$ by $f(H) = f(\Lambda)$, as this allows the Haar measure to be integrated over, leaving only a factor of the volume of $O(n)$, given by Eq. \eqref{eq:onvol}.

We have checked that this formula gives the correct result for integrating over $2 \times 2$ matrices and $3 \times 3$ matrices by using an explicit Euler angle parametrization. For example, the result for $n = 3$ when $f(H)$ depends only on the eigenvalues of $H$ is given by
\begin{align}
\int dH \, f(H) = 2 \pi^2 \int_{-\infty}^\infty d\lambda_1 \int_{-\infty}^{\lambda_1} d\lambda_2 \int_{-\infty}^{\lambda_2} d\lambda_3 \, (\lambda_1 - \lambda_2)(\lambda_1 - \lambda_3)(\lambda_2 - \lambda_3) f(\lambda_1, \lambda_2, \lambda_3).
\end{align}
where we have used $\mathrm{Vol}[O(3)] = 16 \pi^2$.

We can simplify Eq. \eqref{eq:matrixint} a little by choosing an explicit ordering of the eigenvalues. Let $\lambda_1 > \ldots > \lambda_n$. This explicit ordering absorbs the $n!$ factor, yielding
\begin{align}
\int dH \, f(H) = \frac{1}{2^n} \int_{O(n)} (R^T dR) \, \int_{-\infty}^\infty d\lambda_1 \int_{-\infty}^{\lambda_1} d\lambda_2 \ldots \int_{-\infty}^{\lambda_{n-1}} d\lambda_n \, \prod_{i < j} (\lambda_i - \lambda_j) f(R \Lambda R^T).
\end{align}

\section{Integrating Over Rectangular Matrices}\label{app:inta}

In this appendix, we construct a method for integrating over rectangular matrices $A$ when the integrand depends only on $A^T A$. This appendix draws heavily from multivariate statistical theory, and we use a number of mathematical results from Muirhead \cite{Muirhead2005}. Any theorems referenced in this appendix point to theorems in this book.

Consider an $n \times m$ matrix $A$ with $m > n$ that appears in an integral as
\begin{align}
I = \int (dA) \, f(M)
\end{align}
where $M = A^T A$, which is an $m \times m$ symmetric matrix with $m (m-1) / 2$ independent components. As $A$ has $nm$ independent components, we would like to integrate out the extra information contained in $A$. The process of doing so is very similar to the derivation of the Wishart statistical distribution. 

Let us decompose $A = H T$, where $T$ is a $m \times m$ upper triangular matrix with positive diagonal entries, and $H$ is an $n \times m$ matrix with $m$ orthonormal columns ($H$ lives in a object known as the ``Stiefel Manifold''). Note that
\begin{align}
H^T H = \mathbbm{1}_{m \times m}
\end{align}
by construction. $A$ can always be decomposed in this manner when $A$ has rank $m$ (Theorem 3.1.4). When integrating over $A$, the set of matrices that have rank less than $m$ has measure zero, and so we are justified in performing this decomposition.

We then have $M = A^T A = T^T T$. Now we invoke Theorem 2.1.14 to arrive at
\begin{align}
(dA) = 2^{-m} (\det M)^{(n-m-1)/2} (dM) (H^T dH)
\end{align}
where $(H^T dH)$ is the Haar measure on the Stiefel manifold, and $(dM)$ is defined using the exterior product bracket operator. Noting that our integrand depends only on $M$, we can integrate over the Stiefel manifold (Theorem 2.1.15), which has volume
\begin{align}
\int (H^T dH) = \frac{2^m \pi^{mn/2}}{\Gamma_m\left(n/2\right)}
\end{align}
where $\Gamma_m$ is the multivariate Gamma function, defined as
\begin{align}
\Gamma_m(x) = \pi^{m(m-1)/4} \prod_{i=1}^m \Gamma\left[x - \frac{1}{2} (i-1)\right].
\end{align}
Together, we then have
\begin{align} \label{eq:inta}
\int (dA) \, f(M) = \frac{\pi^{mn/2}}{\Gamma_m\left(n/2\right)} \int (dM) \, (\det M)^{(n-m-1)/2} f(M).
\end{align}
Note that in Eq. \eqref{eq:inta}, the integrals over $M$ are only over positive definite matrices. 

To facilitate performing these integrals, we write $M = T^T T$. For concreteness, we work with $m=3$. We parametrize $T$ as
\begin{align}
T = \mat{
	a & d & f \\
	0 & b & e \\
	0 & 0 & c
}.
\end{align}
In order to integrate over $(dT)$ instead of $(dM)$, we use Theorem 2.1.9, which yields
\begin{align}
(dM) = 2^3 a^3 b^2 c \, (dT).
\end{align}
Computing $M$ explicitly, we obtain
\begin{align}
M = T^T T = \mat{
	a^2 & ad & af \\
	ad & b^2 + d^2 & be + df \\
	af & be + df & c^2 + e^2 + f^2
}
\end{align}
with
\begin{align}
\tr{M} = a^2 + b^2 + c^2 + d^2 + e^2 + f^2
\end{align}
and
\begin{align}
\det(M) = a^2 b^2 c^2.
\end{align}
We can then write
\begin{align} \label{eq:intA}
\int (dA) \, f(M) = 
\frac{8 \pi^{3n/2}}{\Gamma_3\left(n/2\right)} \int (dT) \, a^3 b^2 c \, (abc)^{n-4} f(M)
\end{align}
where
\begin{align}
\int (dT) = \int_0^\infty da \int_0^\infty db \int_0^\infty dc \int_{-\infty}^{\infty} dd \int_{-\infty}^{\infty} de \int_{-\infty}^{\infty} df.
\end{align}

\end{document}